\documentclass[review]{elsarticle}

\usepackage{hyperref}

\usepackage{times}
\usepackage{epsfig}
\usepackage{graphicx} 
\usepackage{amsmath} 
\usepackage{amssymb}
\usepackage{algorithm} 
\usepackage{algcompatible}
\usepackage{url}   
\usepackage{tabularx}     
 
\DeclareMathAlphabet{\mathcal}{OMS}{cmsy}{m}{n}
  
\usepackage{amsthm}
\usepackage{bm} 
\usepackage{amsfonts}
\usepackage{latexsym}  
\usepackage{multirow}
\usepackage{subfigure}

\usepackage[T1]{fontenc} 
\usepackage{graphicx} 
\usepackage{mathptmx} 
\usepackage{subfigure}
\usepackage{fancyhdr}
\usepackage{float}
\usepackage{color} 
\usepackage{hhline}
\usepackage{epstopdf}
\usepackage{ctable}
\usepackage[percent]{overpic}

\usepackage{rotating}

\newcommand{\x}{\textbf{x}}
\newcommand{\y}{\textbf{y}}

\newcommand{\bu}{\textbf{u}}
\newcommand{\bv}{\textbf{v}}

\newcommand{\X}{\textbf{X}}

\usepackage{calligra}
\DeclareMathAlphabet{\mathcalligra}{T1}{calligra}{m}{n}

\def\Id{{\textbf{Id}}}

\definecolor{hotpink}{rgb}{1.0, 0.41, 0.71}

\newcommand{\acro}{GRAMIS}

\newcommand{\Real}{\mathbb{R}}

\newcommand{\bSigma}{\bm{\Sigma}}

\newcommand{\normalized}{\widetilde}

\newcommand{\bnu}{\bm \nu}
\newcommand{\bxi}{\bm \xi}
\newcommand{\bmu}{\bm \mu}

\newcommand{\E}{\mathrm{E}}
\newcommand{\Var}{\mathrm{Var}}

\algnewcommand\INITIALIZATION{\item[\textbf{Initialization.}]}%
\algnewcommand\ITERATION{\item[\textbf{Iterative steps.}]}%

\journal{Journal of The Franklin Institute}

\begin{document}

\begin{frontmatter}

\title{Gradient-based Adaptive Importance Samplers}


\author[mymainaddress]{V\'ictor Elvira\corref{mycorrespondingauthor}}
\ead{victor.elvira@ed.ac.uk}
\cortext[mycorrespondingauthor]{Corresponding author}

\author[mysecondaryaddress]{\'Emilie Chouzenoux}
\author[denizlabel]{\"Omer Deniz Akyildiz}
\author[lucalabel]{Luca Martino}

\address[mymainaddress]{School of Mathematics, University of Edinburgh, UK.}
\address[mysecondaryaddress]{CVN, INRIA Saclay, CentraleSupélec, France.}
\address[denizlabel]{Dept.~of Mathematics, Imperial College London,  UK.}
\address[lucalabel]{Universidad Rey Juan Carlos, Spain.}

\begin{abstract}
Importance sampling (IS) is a powerful Monte Carlo methodology for the approximation of intractable integrals, very often involving a target probability distribution. The performance of IS heavily depends on the appropriate selection of the proposal distributions where the samples are simulated from. In this paper, we propose an adaptive importance sampler, called \acro, that iteratively improves the set of proposals. The algorithm exploits geometric information of the target to adapt the location and scale parameters of those proposals. Moreover, in order to allow for a cooperative adaptation, a repulsion term is introduced that favors a coordinated exploration of the state space. This translates into a more diverse exploration and a better approximation of the target via the mixture of proposals. Moreover, we provide a theoretical justification of the repulsion term. We show the good performance of \acro~in two problems where the target has a challenging shape and cannot be easily approximated by a standard uni-modal proposal. 
\end{abstract}


\end{frontmatter}


\section{Introduction}
\label{sec:intro}

The approximation of intractable integrals is a common statistical task in many applications of science and engineering.
A relevant example is the case of Bayesian inference arising for instance in statistical machine learning. A posterior distribution of unknown parameters is constructed by combining data (through a likelihood model) and previous information (through the prior distribution). 
Unfortunately, the posterior is often unavailable, typically due to the intractability of the marginal likelihood.

Monte Carlo methods have proved to be effective in those relevant problems, producing a set of random samples that can be used to approximate a target probability distribution and integrals related to it \cite{Robert04,Liu04b,Owen13}.
Importance sampling (IS) is one of the main Monte Carlo families, with solid theoretical guarantees \cite{Owen13,elvira2021advances}. 
The vanilla version of IS simulates samples from the so-called proposal distribution. Then, each sample receives an associated weight which is computed by taking into account the mismatch between this proposal probability density function (pdf) and the target pdf.
%
%
While the IS mechanism is valid under very mild assumptions \cite{Owen13,elvira2021advances}, the efficiency of the estimators is driven by the choice of the proposal distribution. 
Unfortunately this choice is a difficult task, even more in contexts where one has access to the evaluation of an unnormalized version of the posterior distribution, as it is the case in Bayesian inference. 
The two main methodological directions to overcome the limitations in IS are the usage of several proposals, which is known as multiple IS (MIS) \cite{elvira2019generalized}, and the iterative adaptation of those proposals, known as adaptive importance sampling (AIS) \cite{bugallo2017adaptive}.
%

There exist a plethora of AIS algorithms, and we refer the interested reader to \cite{bugallo2017adaptive}. There are also strong theoretical guarantees for some subclasses of AIS algorithms, see, e.g., \cite{Douc07a, akyildiz2021convergence, akyildiz2022global}.
These AIS methods can be arguably divided in three main categories. The first category is based on sequential moment matching and includes algorithms that implement Rao-Blackwellization in the temporal estimators (\cite{CORNUET12,marin2019}), extend to the MIS setting \cite{APIS14}, or are based on a sequence of transformations that can be interpreted as a change of proposal \cite{paananen2021implicitly}. 
A second AIS family comprises the population Monte Carlo (PMC) methods which use resampling mechanisms to adapt the proposals \cite{Douc05,li2015resampling}.
The PMC framework was first introduced in \cite{Cappe04} and then extended by the incorporation of stochastic expectation-maximization mechanisms \cite{Cappe08}, clipping of the importance weights \cite{koblents2013robust,martino2018comparison}, improving the weighting and resampling mechanisms \cite{elvira2017improving,elvira2017population}, targeting the estimation of rare-event probabilities \cite{miller2021rare}, or introducing optimization schemes \cite{elvira2022optimized}.

Finally, a third category contains AIS methods with a hierarchical or layered structure. Examples of these algorithms are those that adapt the location parameters of the proposals using a Markov chain Monte Carlo (MCMC) mechanism \cite{martino2017layered,schuster2018markov,rudolf2020metropolis,mousavi2021hamiltonian}. 
In this category, we also include methods that exploit geometric information about the target for the adaptation of the location parameters, yielding to optimization-based adaptation schemes. 
In the layered mechanism, the past samples do not affect the proposal adaptation which is rather driven by the geometric properties of the targeted density. However, there also exists hybrid mechanisms, e.g., the O-PMC framework which implements resampling and also incorporates geometric information \cite{elvira2022optimized}. 

%
%
%
\subsection{Contribution within the state of the art}

In this paper, we propose the gradient-based adaptive multiple importance sampling (\acro) method, which falls within the layered family of AIS algorithms. Its main feature is the exploitation of geometric information about the target by incorporating an optimization approach. 
It has been shown that geometric-based optimization mechanisms improve the convergence rate in MCMC algorithms (see the review in \cite{Pereyra16}) and in AIS methods (e.g., \cite{elvira2022optimized}). 
In the context of MCMC, the methods in \cite{roberts2002langevin,durmus2016efficient,Schreck16} are called Metropolis adjusted Langevin algorithms (MALA). The Langevin-based adaptation included in their MCMC proposal updates reads as a noisy gradient descent (called drift term) that favors the exploration of the target in areas with larger probability mass, resulting in a larger acceptance probability in the Metropolis-Hastings (MH) step. Preconditioning can be added for a further improvement of the exploration. To do so, local curvature information about the target density is used to build a matrix scaling the drift term. Fisher metric \cite{Fisher_MCMC_1}, Hessian matrix \cite{Newton_MCMC_1,Newton_MCMC_2,qi2002hessian}, or a tractable approximation of it~\cite{marnissi2018,Marnissi2014Eusipco,MarnissiEntropy2018,Simsekli2016} have been considered for that purpose. Within AIS, the algorithms in \cite{schuster2015gradient,fasiolo2018langevin} adapt the location parameters via several steps of the unadjusted Langevin algorithm (ULA) \cite{Roberts_MALA}. {GRAMIS also connects with the Stein variational gradient descent (SVGD) algorithm \cite{gallego2018stochastic} in the spirit of enhancing the exploratory behavior of the adaptive algorithm. One difference is that SVGD is an MCMC algorithm, while GRAMIS belongs to the family of layered AIS algorithms. Thus, compared to SVGD, GRAMIS requires less stringent convergence guarantees in the adaptive process for its IS estimators to be consistent. Another difference is that the repulsion term in SVGD builds upon reproducing kernel Hilbert space (RKHS) arguments, while in GRAMIS we justify the repulsion by connecting it to Poisson fields, which also translates into a different adaptive scheme.}


A limitation that is present in most AIS algorithms is the lack of adaptation of the scale parameters of the proposals, e.g., the covariance matrices in case of Gaussian proposals. 
However, suitable scale parameters are essential for an adequate performance of the AIS algorithms, and the alternative to their adaptation is setting them a priori, which is in general a difficult task.  
The inefficiency of AIS algorithms that do not implement adaptation in the scale parameters is particularly damaging in high dimensions and where the target presents strong correlations that are unknown a priori.
%
A covariance adaptation has been explored via robust moment-matching mechanisms in \cite{el2018robust,el2019recursive}, second-order information in \cite{fasiolo2018langevin,elvira2022optimized}, and sample autocorrelation \cite{schuster2015gradient}.
The proposed \acro~algorithm implements an adaptation of the covariance by using second-order information (when it yields to a definite positive matrix). 
%
In particular, the covariance adaptation of each proposal is adapted by using the Hessian of the logarithm of the target, evaluated at the updated location parameter. The second-order information is also used to pre-condition the gradient in the adaptation of the location parameters. 

Another limitation in AIS algorithms is the lack of cooperation (or insufficient cooperation) between the multiple proposals at the adaptation stage. 
Some of the few algorithms that implement a type of cooperation can be found in \cite{Cappe08} through a probabilistic clustering of all samples, and in \cite{elvira2017improving} through a resampling that use the deterministic mixture weights.
In the paper, we implement an explicit repulsion between the proposals during the adaptation stage in order to improve the exploration of the algorithm. 

Finally, \acro~implements the balance-heuristic estimator (also called deterministic mixture) importance weights  \cite{Veach95,Owen00,sbert2022generalizing}, which have been shown a theoretical superiority (in the unnormalized IS estimators) \cite{elvira2019generalized} and a superior performance in other types of AIS algorithms (e.g., in \cite{APIS15,elvira2017improving,elvira2022optimized}).\footnote{{The GRAMIS algorithm builds upon the GAPIS algorithm, which was presented in the conference paper \cite{elvira2015gradient}. In GRAMIS, on top of updating the location and scale parameters of the proposal parameters by incorporating a repulsion term and using the first and second-order information of the target, we go beyond our preliminary work in the three following ways. First, we pre-condition the gradient steps in the adaptation of the location parameters by using second-order information. In this way, we have a much more efficient adaptation, and we reduce a hyper-parameter w.r.t. the GAPIS algorithm. Second, we find a strong connection with Poisson fields, which supports the theoretical justification of the algorithm and opens the door for further analysis and methodological improvements.  Finally, based on this connection, we crucially modify the repulsion term between proposals (e.g., exponentiating the distance between proposal to the dimension of the latent space), which allows the method to scale in dimensions.}}

\subsection{Notation}
We denote by $\|\x\|={\sqrt{\x^\top \x}}$ the Euclidean norm of $\x \in \Real^d$, where $\top$ is the transpose operation and $\Real^d$ is the $d$-dimensional Euclidean space. $\nabla$ and $\nabla^2$ denote the first and second differential operators in $\x$, i.e., resulting in the gradient vector and the Hessian matrix, respectively. The operator $\bSigma \succ 0$ is used to describe $\bSigma$ as a positive definite matrix.  
$\Id_{d}$ is the identity matrix of dimension $d$. Bold symbols are used for vectors and matrices. We use the $p$ notation for probability density functions (pdf) only when it is unambiguous. 

\subsection{Structure of the paper}

The rest of the paper is structured as follows.
Section \ref{sec_back} introduces the problem and relevant background.
In Section \ref{sec_proposed}, we describe the \acro~algorithm. 
We provide numerical examples in \mbox{Section \ref{sec_results}}.
Section \mbox{\ref{sec_conclusion}} closes the paper with some conclusion.

\section{Background in importance sampling}
\label{sec_back}

{We motivate importance sampling (IS) by considering the Bayesian inference setup, where the goal is characterizing the posterior distribution}
\begin{equation}
	\normalized{\pi}(\x| \y)
		= \frac{\ell(\y|\x) {\pi_0}(\x)}{Z(\y)},  
\label{eq_posterior}
\end{equation}
{where
	 $\x\in\Real^{d_x}$ is the variable associated to the r.v. $\X$ of the vector of unknowns to be estimated;
	 $\y\in\Real^{d_y}$ represents the available data;
	 $\ell(\y|\x)$ is the likelihood function; and
	 ${\pi_0}(\x)$ is the prior distribution.}

We consider the problem where one must compute the integral
\begin{equation}
	I =  \int h(\x) \normalized{\pi}(\x|{\y}) d\x = \frac{1}{Z(\y)} \int h(\x) \pi(\x|{\y}) d\x,
\label{eq_integral}
\end{equation}%
where $h$ is any integrable function w.r.t. $\widetilde \pi(\x|{\y})$. Such problem can arise for instance in the field of Bayesian learning, when $\y$ gathers the available data to train a model described by vector $\x$~\cite{welling,Rasmussen}. 

{In many cases of interest, the integral in Eq. \eqref{eq_integral} does not have an analytic form. This happens very often in Bayesian inference when the marginal likelihood, $Z(\y) \triangleq \int  \pi(\x|\y) d\x$, is intractable. 
In these cases, one has access only to the non-negative function $\pi(\x|\y)\triangleq\ell(\y|\x) {\pi_0}(\x) =  Z(\y) \normalized\pi(\x| \y)$. }
{In the rest of the paper, we alleviate the notation by denoting $Z$, $\pi(\x)$, and $\widetilde \pi(\x)$, i.e., dropping $\y$ from the notation, since we aim at targeting more generic setups beyond the Bayesian inference problem.}  
%

\begin{table} [h!]
{
{
\setlength{\tabcolsep}{2pt}
\def\marginwidth{1.5mm}
\begin{center}
\begin{tabularx}{\textwidth}{|l|X|l|}
\hline
    \textbf{Notation} & \textbf{Description}  \\
    \hline                             
\hline
 $\x\in\Real^{d_x}$ & variable associated to the r.v. $\X$ of interest \\
\hline
 $\y\in\Real^{d_y}$  & vector of data \\
\hline
$\widetilde\pi(\x) \triangleq \widetilde \pi(\x|\y)$ &   target pdf \\
\hline
  $\pi(\x) \triangleq \pi(\x|\y) \triangleq\ell(\y|\x) {\pi_0}(\x) =  Z(\y) \normalized\pi(\x| \y)$   &   unnormalized target density function\\
\hline
  $Z \triangleq  Z(\y)$   &  normalizing constant \\
\hline                                              
  $h(\x)$  & test function \\
\hline                        
	 $\ell(\y|\x)$  & likelihood function   \\
	\hline
	 ${\pi_0}(\x)$ &  prior pdf\\
	\hline
 	$\widehat I$ &  unnormalized IS (UIS) estimator \\
 		\hline
 	$\widetilde I$ &  self-normalized IS (SNIS) estimator \\
	\hline
	$N$ &  number of proposals\\
	\hline
	$K$ &  number of samples per proposal\\
	\hline
	$T$ &  number of iterations\\
		\hline
		$q_{n}^{(t)}(\x;\bmu_n^{(t)},\bSigma_n^{(t)},\bxi_{n})$ &  $n$-th proposal pdf at $t$-th iteration\\
		\hline
		$ \bmu_n^{(t)} $ & location parameter of the $n$-th proposal pdf at $t$-th iteration\\
		\hline
		$ \bSigma_n^{(t)} $ & scale parameter of the $n$-th proposal pdf at $t$-th iteration\\
		\hline
		$ \bxi_n  $ & non-adapted parameters of the $n$-th proposal pdf\\
	\hline
		$\x_{n,k}^{(t)}$ & $k$-th sample from the $n$-th proposal at $t$-th iteration\\
	\hline
	$w_{n,k}^{(t)}$ & importance weight associated to $\x_{n,k}^{(t)}$\\
	\hline
\end{tabularx}
\end{center}
\caption{Summary of notation of the main variables, functions, constants, and parameters. Some functions are explicit for a generic Bayesian inference problem.
}
\label{table_banana_gramis}
}
}
\end{table}

\subsection{Importance sampling}
\label{sec_IS}

Importance sampling (IS) is one of the main Monte Carlo methodologies for the approximation of distributions and related integrals. 
In IS, $K$ samples $\x_k$ are simulated from an alternative distribution $q(\x)$, called proposal, and receive {an} importance weight $w_k$. The procedure comprises two basic steps: 

\begin{enumerate}
	\item \textbf{Sampling.} $K$ samples are drawn as
	$$\x_k \sim q(\x), \qquad k=1,...,K.$$ 
	\item \textbf{Weighting.} Each sample is associated to an importance sampling
	$$w_k = \frac{\pi(\x_k)}{q(\x_k)}, \qquad k=1,...,K.$$
\end{enumerate}
The resulting sets of samples $\{ \x_k\}_{k=1}^K$ and weights $\{ w_k\}_{k=1}^K$ are used in order to produce estimators that approximate $I$ in Eq. \eqref{eq_integral}. When $Z$ is available, it is possible to produce the unnormalized IS (UIS) estimator
	\begin{equation}
	\widehat I = \frac{1}{KZ} \sum_{k=1}^{K} w_k h(\x_k).
	\label{eq_UIS}
	\end{equation}
	If $Z$ is not available, the alternative is to use the self-normalized IS (SNIS) estimator,
\begin{equation}
\widetilde I = \sum_{k=1}^{K} \overline  w_k h(\x_k),
\label{eq_SNIS}
\end{equation}
where $\overline w_k = w_k/\sum_{j=1}^K w_j$, $k=1,...,K$ are the importance weights. 

The UIS estimator is unbiased and consistent. The SNIS estimator is consistent and has a bias which vanishes at a faster rate than the variance when $K$ grows. 
%
The optimal proposal of the UIS estimator is $q(\x) \propto |h(\x)|\pi(\x)$ \cite{Robert04,Liu04b}, while the optimal proposal of the SNIS estimator is (approximately) $q(\x) \propto |h(\x)|\pi(\x)$ \cite{Owen13}.   


\subsection{Multiple importance sampling}
\label{sec_mis}

One of the most common strategies is to use several proposals, $\{q_n(\x)\}_{n=1}^N$ \cite{Hesterberg95,Veach95,Owen13}. The last years have witnessed and increased of attention in MIS \cite{kondapaneni2019optimal,sbert2018multiple,elvira2015efficient,elvira2016heretical,elvira2016multiple} (see a generic framework with theoretical analysis in \cite{elvira2019generalized}). 
%
%
It has been shown that several weighting and sampling schemes are possible, i.e., that lead to consistent UIS and SNIS estimators \cite{elvira2019generalized}. We consider the simplified example where we simulate $K=N$ samples from the set of proposals. Then, one possibility is to simulate exactly one sample per proposal as $\x_n \sim q_n(\x)$, $n=1,...,N$. Then, two popular weighting approaches are 
%
the standard MIS (s-MIS), 
$$w_n= \frac{\pi(\x_n)}{{q_n(\x_n)}}, \quad n=1,\ldots, N,$$
and the deterministic mixture MIS (DM-MIS),
\begin{equation} 
	w_n=\frac{\pi(\x_n)}{\frac{1}{N}\sum_{j=1}^{N}q_j(\x_n)}=\frac{\pi(\x_n)}{\frac{1}{N}\sum_{j=1}^{N}q_j(\x_n)}, \quad n=1,\ldots, N.\nonumber
\label{f_dm_weights_static}
\end{equation}
In \cite{elvira2019generalized}, it is proved that DM-MIS weights provide an UIS estimator with less variance compared to s-MIS, for any function $h$, any target $\widetilde \pi$, and any set of proposals. 

\subsection{Optimization-based samplers}

The performance of sampling algorithms depends greatly on the choice of the proposal distribution. Proposals parametrized with static parameters are easier to implement and manipulate, as they require minimal self tuning, but this simplicity comes at the price of a suboptimal (since not adaptative) target exploration. To cope with this issue, several methods have been proposed to iteratively update the proposal, along the sampling algorithm iterations, so as to improve and accelerate the target exploration. The most common technique is to resort to a Langevin-based approach, where gradient descent steps (assuming differentiability of $\log \pi$) are performed to adapt the proposal mean (i.e., location). The discretization of the  Langevin dynamics leads to the unadjusted Langevin algorithm (ULA) \cite{Roberts_MALA}, which can also be viewed as a gradient descent algorithm perturbed with an independent and identically distributed (i.i.d.) stochastic error. The convergence of ULA is discussed in~\cite{Talay1990,Roberts_MALA}. However, in most situations, the stationary distribution of the samples produced by ULA differs from the target $\pi$~\cite{Durmus2019}, due to the discretization of the Langevin dynamic. MALA~\cite{roberts2002langevin} tackles this issue by introducing an MH strategy, hence guaranteeing ergodic convergence to the sought target law. Accelerated variants of MALA have been investigated, based on preconditioning techniques to account for more information (e.g., curvature) about the target~\cite{marnissi2018,Fisher_MCMC_1,Fisher_MCMC_2,qi2002hessian,XIFARA201414,roberts2002langevin,Sabanis2018}. For instance, the Newton MH strategy \cite{Fisher_MCMC_2,qi2002hessian} consists in combining an MH procedure with a stochastic Newton update involving
the inverse (or an approximation of it, when undefined or too complex) of the Hessian matrix of $\log \pi(\x)$.
This approach will serve as starting point for introducing a proposal adaptation within our novel approach \acro. 

\section{The \acro~algorithm} 
\label{sec_proposed}

We now describe \acro, the proposed AIS algorithm, in Table \ref{alg_gapis}. The algorithm runs over $T$ iterations, adapting $N$ proposals, and simulating $K$ sampler per proposal and iteration. {At the $t$-th iteration, we adapt the location parameter, $\bmu_n^{(t)}$, and the scale parameter, $\bSigma_n^{(t)}$, of each proposal, with $n=1,...,N$. The non-adapted parameters are denoted as $\bxi_n$ (for instance, the degrees of freedom when using Student's t-proposals).}

First,  the location parameters are updated in \eqref{mu_adapt} following a gradient step that includes an optimized stepsize $\theta_n^{(t-1)}$, and first and second-order information of the log-target at the previous location parameter $\bmu_n^{(t-1)}$ of each proposal (see Section \ref{sec_location} for more details). A repulsion term between each pair of proposals (i.e., between $j$-th and $i$-th proposals), $ {\bf r}_{i,j}^{(t-1)}$, is introduced. This repulsion force  which inversely proportional to the  (Euclidean) distance $||{\bf d}_{i,j}||  = ||\bmu_{i}^{(t-1)} - \bmu_{j}^{(t-1)}||$. More practical information about the choice of repulsion strategy can be found in Section \ref{sec_repulsion}. Second, the scale parameters are updated by using second-order information of the log-target (see Section \ref{sec_scale}). Third, $K$ samples are simulated from each proposal. The importance weights are computed in Eq. \eqref{eq:ISweights}. {Note that we implement the DM-MIS weights as in Eq. \eqref{f_dm_weights_static}, which is guaranteed to reduce the variance of the UIS estimators \cite[Theorems 1 and 2]{elvira2019generalized}.}
%
%
\acro~returns $KNT$ weighted samples that can be used to estimate both $I$ and $Z$ (in the case it is unknown). The simplest version of those estimators is given below.   

\begin{itemize}
	\item UIS estimator:
		\begin{equation}
	\widehat I = \frac{1}{KTNZ} \sum_{t=1}^{T}\sum_{n=1}^{N}\sum_{k=1}^{K} w_{n,k}^{(t)} h(\x_{{n},k}^{(t)}).
	\label{eq_UIS_AIS}
	\end{equation}
	\item SNIS estimator:
	\begin{equation}
\widetilde I =  \sum_{t=1}^{T}\sum_{n=1}^{N}\sum_{k=1}^{K} \widetilde w_{n,k}^{(t)} h(\x_{{n},k}^{(t)}),
\label{eq_SNIS_AIS}
\end{equation}
where 
\begin{equation}
\widetilde w_{n,k}^{(t)}  =   \frac{w_{n,k}^{(t)} h(\x_{{n},k}^{(t)})}{\sum_{t=1}^{T}\sum_{n=1}^{N}\sum_{k=1}^{K} w_{n,k}^{(t)}}
\end{equation}
are the re-normalized weights.
\item estimator of $Z$:
\begin{equation}
	\widehat Z = \frac{1}{KTNZ} \sum_{t=1}^{T}\sum_{n=1}^{N}\sum_{k=1}^{K} w_{n,k}^{(t)}.
	\label{eq_Zhat_AIS}
	\end{equation}
\end{itemize}

\begin{table}[!t]
\centering
{
	\caption{{{\color{black} \acro.}}}
	
	\begin{tabular}{|p{0.95\columnwidth}|}
    \hline
		\footnotesize
		\begin{enumerate}
			\item {\bf [Initialization]}:  Initialize the proposal means $\bmu_{n}^{(0)}$ and the non-adapted parameters   $\bxi_{n}$. 
			Compute the scale parameter matrix $\bSigma_n^{(0)}$ using \eqref{eq:covadapt1b}.
			\vspace*{6pt}
			\item {\bf[For $\bm t \bm= \bm 1$ to  $\bm T$]}: 
			\begin{enumerate}
							\item {\bf Mean adaptation:}
				\begin{enumerate}
\item Compute the stepsize $\theta_n^{(t-1)}$ using the backtracking procedure so as to satisfy \eqref{eq:backtrack}.
\item The mean of the $n$-th proposal is adapted as
	\begin{equation}	\bmu_{n}^{(t)} = \bmu_{n}^{(t-1)}  + \theta_n^{(t-1)} \bSigma_n^{(t-1)} \nabla\log\left(\pi(\bmu_{n}^{(t-1)})\right) + \sum_{j=1,j\neq n}^{N} {\bf r}_{n,j}^{(t-1)},
\label{mu_adapt}
	\end{equation}
with 
	\begin{equation}
	  {\bf r}_{n,j}^{(t-1)} = G_t  \frac{m_nm_j}{\lVert {\bf d}_{n,j}^{(t-1)} \rVert^{d_x}}{\bf d}_{n,j}^{(t-1)},
\label{repulsion}
	\end{equation}
{where $\lVert \cdot \lVert$ represents the norm operator, ${\bf d}_{n,j}^{(t-1)}  = \bmu_{n}^{(t-1)} - \bmu_{j}^{(t-1)}$, and $m_n, m_j>0$ are two positive terms that depend on the $n$-th and $j$-th proposals respectively.} 
\end{enumerate}

\item {\bf Covariance adaptation:} The covariance matrix of the $n$-th proposal $\bSigma_n^{(t)}$ is adapted using \eqref{eq:covadapt1b}.

\item {\bf Sampling steps:}
\begin{enumerate}
	\item Draw $K$ independent samples from each proposal, i.e., $\x_{n,k}^{(t)} \sim q_{n}^{(t)}({\bf x};\bmu_{n}^{(t)},\bSigma_n^{(t)},\bxi_{n})$ for $k=1\ldots,K$  and $n=1,\ldots,N$.
	\item Compute the importance weights,
		\begin{equation}
			w_{n,k}^{(t)} = \frac{\pi(\x_{n,k}^{(t)})}{\frac{1}{N}\sum_{j=1}^N q_{j}^{(t)}(\x_{n,k}^{(t)};\bmu_{n}^{(t)},\bSigma_n^{(t)})}, 
		\label{eq:ISweights}
		\end{equation}
		for $\quad  n=1,\ldots,N$, and  $k=1,\ldots,K$.
\end{enumerate}

			\end{enumerate}
					\item {\bf [Output]}: 
				Return the pairs $\{\x_{n,k}^{(t)}, {w}_{n,k}^{(t)}\}$, for
				$n=1,\ldots,N$, $k=1,\ldots,K$, and $t=1,\ldots,T$.
			\end{enumerate}\\
			\hline
		\end{tabular}\label{alg_gapis}
	}
\end{table}

\subsection{Adaptation of location parameters}
\label{sec_location}

The location parameters are adapted as in Eq. \eqref{mu_adapt}. The adaptation process implements a Newton ascent on $\log \pi$. The gradient of the log target is evaluated at the previous location parameter and pre-conditioned by $\bSigma_n^{(t-1)}$ where $\bSigma_n^{(t-1)}$ is the same that we use in the previous section in order to update the covariance.
We furthermore introduced $\theta_n^{(t-1)}\in (0,1]$, which is a stepsize tuned according to a backtracking scheme in order to avoid the degeneracy of the Newton iteration, and thus of our adaptation scheme, for non log-concave distributions. Starting with unit stepsize value, we reduce it by factor $1/2$ until the condition below is met: 
\begin{equation}
\pi\left(\bmu_{n}^{(t-1)}  + \theta_n^{(t-1)} \bSigma_n^{(t-1)} \nabla\log\left(\pi(\bmu_{n}^{(t-1)})\right)\right) \geq
\pi \left(\bmu_{n}^{(t-1)}\right),
\label{eq:backtrack} 
\end{equation}
{i.e., the repulsion term is not considered in the backtracking scheme.}


The update in Eq. \eqref{mu_adapt} also incorporates an innovative repulsion term among proposals. The purpose is to efficiently explore the space in a cooperative manner. This repulsion term admits several interpretations. It can be seen as an over-spreading of the mixture proposal, i.e., a safer choice of mixture that will overweight the tails of the target \cite{Owen00}. Also, it can be interpreted as a negative coupling among proposals. It shares connections with MCMC algorithms that implement interacting parallel chains, with a similar spirit as it is done in MCMC \cite{martino2014orthogonal,martino2015smelly}. In Section \ref{sec_repulsion}, we discuss the practical repulsion schemes, and the rationale of the adaptation is discussed in Section~\ref{sec_interpretation}.

\subsection{Adaptation of scale parameters}
\label{sec_scale}
We implement a Newton-based strategy, {inherited from the literature of optimization \cite{Nocedal}}, to exploit the Hessian of $\log \pi$ in the update of the scale parameter. In general scenarios, the convexity of $- \log \pi$ is not ensured, and numerical issues might arise when computing the inverse of its Hessian. We thus propose to introduce a safe rule in our adaptation method, so that
\begin{equation}
\label{eq:covadapt1b}
\bSigma_n^{(t)} = 
\begin{cases}
\left(- \nabla^2 \log \pi({\bmu}_n^{(t)})\right)^{-1}, & \text{if} \; \nabla^2 \log \pi({\bmu}_n^{(t)})\succ 0,\\
\bSigma_n^{(t-1)}, & \text{otherwise}.
\end{cases}
\end{equation}
The scaling matrix thus incorporates second order information on the target, whenever this yields to a definite positive matrix. Otherwise, it inherits the covariance of the proposal of the previous iteration, where $\bSigma_n^{(0)}$ is set to a predefined default value (typically, a scalar times the identity matrix).

\subsection{Design of the repulsion scheme}
\label{sec_repulsion}

%
%
Our repulsion term is parameterized by a common time-dependent constant $G_t$ and a proposal-dependent constant, $m_n$, for each $n=1,\ldots,N$. By construction, Eq. \eqref{repulsion} implies that the repulsion term vanishes whenever the proposals get further away (in Euclidean distance). 

The interpretation of the functional form in Eq. \eqref{repulsion} is discussed in the next section. 
The simpler choice is to keep $m_n=1$, for all $n=1,\ldots,N$, and to fix the common term $G_t$ to be constant over the iterations, i.e., $G_t=G$. 
In this case, the repulsion never vanishes with the consequence of leading to a potential equilibrium positioning of the proposals in such a way that the interpreted mixture proposal, $\widetilde q^{(t)}(\x)\triangleq \frac{1}{N}\sum_{n=1}^N q_n^{(t)}(\x;\bmu_n^{(t)},\bSigma_n^{(t)},\bxi_{n})$   would overweight the tails of the target distribution. 
In this case, it is not guaranteed that the proposal adaptation converges in finite $t$. 
%
An alternative is to reduce the repulsion term in such a way that $r_{n,j}^{(t)} \to 0$ when $t\to \infty$. 
A natural choice is a decaying term in the form of 
\begin{equation}
\label{EqG1}
G_t=\exp\left(-\beta t\right), \quad \quad \beta>0.
\end{equation}
In such case, if the Newton scheme converges to a local maximum for each proposal, the whole mixture approximation would converge to a mixture of local Laplace approximations.
The choice of $\beta$ can be easily set depending on the repulsion strength desired in the last iteration, e.g., a $1\%$ of attenuation in the last iteration leads to $\beta = \frac{-\log(0.01)}{T-1}$. It is also possible to set the repulsion term to zero in the last iteration, so a final set of samples can be simulated. 
%
%

\subsection{{Interpretation of the repulsion term}}
\label{sec_interpretation}
We can interpret the repulsion term of Eq. \eqref{repulsion} in general physical terms. The following discussion uses the particle interpretation of the repulsion term mentioned in \cite{elvira2015gradient}, formalizing it using the notion of Poisson fields \cite{xu2022poisson}. {In what follows, we first introduce the notion of Poisson fields in Section~\ref{sec:Poisson_sec} and then connect this to our algorithm in Section~\ref{sec:repulsion_as_poisson}.}
\subsubsection{{Poisson fields}}\label{sec:Poisson_sec}
This section summarizes Poisson fields as in \cite{xu2022poisson}. Let $\rho(\mathbf{v}): \mathbb{R}^{d_x} \to \mathbb{R}$ be a \textit{source function} with a compact support. In this setting, the Poisson equation is defined as
\begin{align}\label{eq:Poisson}
\nabla^2 \varphi(\mathbf{v}) = -\rho(\mathbf{v}),
\end{align}
where $\varphi(\bv):\mathbb{R}^{d_x} \to \mathbb{R}$ is called \textit{the potential function}. The gradient $\nabla \varphi(\bv)$ is usually interpreted as a \textit{field} and may have a physical meaning analogous to the low dimensional cases. For example, writing $\mathbf{E}(\bv) = -\nabla \varphi(\bv)$ converts the Poisson equation into $\nabla \cdot \mathbf{E} = \rho$ which is Gauss' law in physics \cite{xu2022poisson}. In this case, $\mathbf{E}$ would be a $d_x$-dimensional generalization of the 3-dimensional electric field. {Similarly $\rho$ would be a $d_x$ dimensional analogue of the electrical charge density.}

{Equation~\eqref{eq:Poisson} has a unique solution (with extra regularity conditions \cite[Lemma~1]{xu2022poisson}) given as
\begin{align*}
\varphi(\bu) = \int G(\bu, \bv) \rho(\bv) \mathrm{d} \bv,
\end{align*}
where
\begin{align*}
G(\bu, \bv) = \frac{1}{(d_x - 2) S_{d_x-1}(1)} \frac{1}{\|\bu-\bv\|^{d_x -2}},
\end{align*}
and 
\begin{align*}
S_{d_x - 1}(1) = \frac{2 \pi^{d_x/2}}{\Gamma(d_x / 2)}
\end{align*}
is a constant equals to the surface area of the unit $(d_x - 1)$ sphere  (note that the particle interpretation is then valid for in $\mathbb{R}^{d_x}$ {for $d_x \geq 3$}). Then, the negative gradient field is called a \textit{Poisson field} \cite{xu2022poisson} given as
\begin{align*}
-\nabla \varphi(\bu) = - \int \nabla_{\bu} G(\bu, \bv) \rho(\bv) \mathrm{d} \bv,
\end{align*}
where
\begin{align*}
\nabla_{\bu} G(\bu, \bv) = -\frac{1}{S_{d_x-1}(1)}\frac{\bu-\bv}{\|\bu-\bv\|^{d_x}}.
\end{align*}
The property of the Poisson field $-\nabla\varphi(\bu)$ is that, it creates a field that moves a particle away from sources $\rho(\bv)$. In the case of a single source (i.e., $\rho(\bv)$ is a Dirac), this quantity would create a repulsive effect w.r.t. this particle. In what follows, we will interpret the repulsion term in our GRAMIS scheme in terms of an \textit{empirical} Poisson field where the source distribution will coincide with the GRAMIS proposals in the previous iteration.}

\subsubsection{{Repulsion term as an empirical Poisson field}}\label{sec:repulsion_as_poisson}

In this section, for the ease of presentation, we consider a simplified version of the mean adaptation defined in \eqref{mu_adapt}. In particular, we consider a fixed, scalar step-size $\theta_n^{(t-1)} = \gamma$ for all $n = 1,\ldots,N$ and $t = 1,\ldots,T$ and $\bSigma_n^{(t-1)} = I_{d_x}$, where $I_{d_x}$ is an identity matrix. {In this case, the update rule takes this form:}
\begin{equation}
\bmu_{n}^{(t)} = \bmu_{n}^{(t-1)}  + \gamma \nabla\log\left(\pi(\bmu_{n}^{(t-1)})\right) + \sum_{j=1,j\neq n}^{N} {\bf r}_{n,j}^{(t-1)},
\label{mu_adapt_2}
\end{equation}
where $\gamma>0$ is a scalar step-size. {For the repulsion term, we also provide some choices that would lead to clarification of the role of the repulsion term. In particular, we set
\begin{align}
G_t = \frac{\gamma}{(N-1) S_{d_x - 1}(1)} 
\end{align}
for all $t = 1,\ldots,T$ and $m_n = 1$ for all $n = 1,\ldots,N$.\footnote{{Note that this is not restrictive, as any choice of $G_t$ can be multiplied and divided by $\frac{(N-1) S_{d_x - 1}(1)}{\gamma}$ to define $\tilde{G}_t = \frac{(N-1) S_{d_x - 1}(1)}{\gamma} G_t$. These constants can also be used to choose the parameters $m_n$.}}} {As a consequence, the repulsion term takes the following form:
	\begin{equation}
	  {\bf r}_{n,j}^{(t-1)} =   \frac{\gamma}{(N-1) S_{d_x - 1} (1)} \frac{1}{\lVert {\bf d}_{n,j}^{(t-1)} \rVert^{d_x}}{\bf d}_{n,j}^{(t-1)}.
\label{repulsion_2}
\end{equation}}
{where ${\bf d}_{n,j}^{(t-1)}  = \bmu_{n}^{(t-1)} - \bmu_{j}^{(t-1)}$. Our aim is now to interpret the repulsion term in \eqref{mu_adapt_2} as a Poisson field which creates a repulsive effect for location parameter $n$ and pushes it away from the other location parameters $\bmu_j^{(t-1)}$, with  $j \in \{1,\ldots,N\} \setminus n$.}

{Our first step is to interpret the last term in \eqref{mu_adapt_2} as an empirical estimate of an integral. In order to do so, recall that with the choice of \eqref{repulsion_2}, the last term of \eqref{mu_adapt_2} can be written as}
{\begin{align}\label{eq:repulsion_rewritten}
\sum_{j=1, j \neq n}^N \mathbf{r}_{n,j}^{(t-1)} = \frac{\gamma}{(N-1)} \sum_{j=1, j \neq n} \frac{1}{S_{d_x-1}(1)} \frac{1}{\lVert {\bf d}_{n,j}^{(t-1)} \rVert^{d_x}}{\bf d}_{n,j}^{(t-1)},
\end{align}}
{where ${\bf d}_{n,j}^{(t-1)}  = \bmu_{n}^{(t-1)} - \bmu_{j}^{(t-1)}$. The sum in the r.h.s. above can now be interpreted as an integral approximation. More precisely, let us consider the empirical measure constructed by the sequence $\{\bmu^{(t-1)}_{n}\}_{n=1}^N$ given by
\begin{align*}
\rho_{t-1, n}^{N}(\mathrm{d} \bu) = \frac{1}{{N-1}}\sum_{j=1,j\neq n}^N \delta_{\bmu_j^{(t-1)}}(\mathrm{d} \bu),
\end{align*}
which is defined for the update of $\bmu_n^{(t-1)}$. Using this empirical measure, we can interpret the repulsion term in {\eqref{eq:repulsion_rewritten}} as 
\begin{align}
\frac{\gamma}{(N-1)} \sum_{j=1, j \neq n} \frac{1}{S_{d_x-1}(1)} \frac{1}{\lVert {\bf d}_{n,j}^{(t-1)} \rVert^{d_x}}{\bf d}_{n,j}^{(t-1)} = \gamma \int g(\bmu_n^{(t-1)}, \bv) \rho_{t-1, n}^N(\mathrm{d} \bv),
\end{align}}
with
\begin{align}\label{eq:g_term}
g(\bu, \bv) = \frac{1}{S_{d_x - 1}(1) \| \bu - \bv \|^{d_x}} (\bu - \bv).
\end{align}
{Finally, embedding this into our update rule \eqref{mu_adapt_2}, we obtain}
\begin{align}\label{eq:mean_update_with_integral}
\bmu_{n}^{(t)} = \bmu_{n}^{(t-1)}  + \gamma \nabla\log\left(\pi(\bmu_{n}^{(t-1)})\right) + \gamma \int g(\bmu_n^{(t-1)}, \bv) \rho_{t-1, n}^N(\mathrm{d} \bv),
\end{align}
This implies that, if we set $g(\bu, \bv) = - \nabla_\bu G(\bu, \bv)$, where
\begin{align*}
G(\bu,\bv) = \frac{1}{(d_x-2) S_{d_x-1}(1)} \frac{1}{\|\bu - \bv\|^{d_x - 2}},
\end{align*}
then the last term in \eqref{eq:mean_update_with_integral} can be interpreted as a gradient \cite{xu2022poisson}, i.e.,
\begin{align*}
- \nabla \varphi_{t-1}^N(\bu) = - \int \nabla_u G(\bu, \bv) \rho_{t-1, n}^N(\mathrm{d} \bv).
\end{align*}
{Finally, going back to the update of the location parameters,} we can then rewrite \eqref{mu_adapt_2} as
\begin{equation}\label{mu_adapt_poisson}
\bmu_{n}^{(t)} = \bmu_{n}^{(t-1)}  + \gamma \nabla\log\left(\pi(\bmu_{n}^{(t-1)})\right) - \gamma \nabla \varphi_{t-1}^N({\bm\mu_n^{(t-1)}}),
\end{equation}
where
\begin{align*}
-\nabla \varphi^N({\bm\mu_n^{(t-1)}}) = \frac{1}{{N-1}}\sum_{j=1,j\neq n}^N \frac{\Gamma(d_x/2)}{2 \pi^{d_x/2}} \frac{\mathbf{d}_{n,j}^{(t-1)}}{\lVert {\bf d}_{n,j}^{(t-1)} \rVert^{d_x}}.
\end{align*}
{One can clearly see that $-\nabla \varphi_{t-1}^N(\bu)$ is a \textit{Poisson field} as defined in Section~\ref{sec:Poisson_sec} with $\rho_{t-1,n}^N$ as the \textit{empirical source distribution}. In other words, the term $-\nabla \varphi_{t-1}^N(\bu)$, as a field, would push a location parameter away from all others in that constructs the empirical distribution $\rho_{t-1,n}^N$. The update \eqref{mu_adapt_poisson} has two effects by pushing the $n$-th location parameter to maximize $\pi$ due to the effect of $\nabla\log\left(\pi(\bmu_{n}^{(t-1)})\right)$, and also by staying away from all other location parameters by the repulsion term term $- \nabla \varphi_{t-1}^N({\bm\mu_n^{(t-1)}})$, i.e. \textit{the empirical Poisson field}. More explicitly,} Eq.~\eqref{mu_adapt_poisson} describes the precise balance achieved by the repulsion term. The term $\nabla \log \pi(\cdot)$ in \eqref{mu_adapt_poisson} pushes the means towards the maximum-a-posteriori (MAP) estimate, which, if implemented alone, would cause all means to converge to the MAP (e.g., in settings where target has a single maximum, i.e., MAP is uniquely defined). The repulsion term creates a potential for means to stay away from each other. More precisely, the last term in \eqref{mu_adapt_poisson} pushes means away from \textit{sources}. In our case, sources for the adaptation of the $n$-th location parameter are the other location parameters $\{\bm \mu_j^{(t-1)}\}_{j=1,j\neq n}^N$. In other words, the addition of the term $-\nabla \varphi^N({\bm\mu_n^{(t-1)}})$ to the gradient flow above creates a repulsive effect, pushing the updated mean away from the location parameters, which effectively spreads out the components in the mixture proposal. This interpretation also holds when we introduce back $m_n$, $n = 1,\ldots,N$ terms, effectively determining the strength of the repulsion for a particular mean. From this viewpoint, $G_t$ can also be seen as the adaptive weight that determines whether the repulsion term should be more or less active. High values of $G_t$ might be useful in the initial exploration phase.

\section{Numerical experiments}
\label{sec_results}
\subsection{{Ablation study with Gaussian mixtures}}
\label{ex_mixture}

 {Let us consider a generic mixture of bivariate Gaussian pdfs as
\begin{equation}
(\forall \x \in \mathbb{R}^2) \quad \widetilde{\pi}({\bf x}) = \frac{1}{L}\sum_{\ell=1}^L \mathcal{N}(\x;{\bf \gamma}_\ell,\bSigma_\ell).
\end{equation}
We start considering a toy example with $L=2$ components to better understand the behavior of \acro. In this case, the means are  ${\bf \gamma}_1=[-5, -5]^{\top}$and ${\bf \gamma}_2=[6, 4]^{\top}$, and the covariance are  $\bSigma_1=[0.25,\ 0;0, \ 0.25]$ and $\bSigma_2=[0.52, \ 0.48; 0.48, \ 0.52]$. We run~\acro~displaying the adaptive behavior of different ablated version of the algorithm. We set $N=50$ Gaussian proposals, $T=20$ iterations, and $K=20$ samples per proposal and iteration. {The location parameters of the proposals are randomly initialized in the square {$[1,6]\times[1,6]$}.}

Figure \ref{fig_toy_2D} shows the final location parameters (black dots) and scale parameters (black ellipses) of the proposals at time $t=20$ for four ablated versions of \acro. Plot (a) shows the modified \acro~without preconditioning matrix in the gradient update (as in \cite{elvira2015gradient} with $\lambda=0.1$) and $G_t=0$ (no repulsion). {In this setting, the adaptation of the location parameter simplifies into
	\begin{equation}	\bmu_{n}^{(t)} = \bmu_{n}^{(t-1)}  + \lambda \nabla\log\left(\pi(\bmu_{n}^{(t-1)})\right).
	\end{equation}
	}
 The arbitrary (suboptimal) step-size delays the convergence of the location parameter to the mode. Plot (b) shows \acro~with $G_t=0$ (no repulsion). The adaptation is effective in recovering one mode, but not in finding the second mode. Plot (c) shows \acro~with constant repulsion $G_t=0.5$. The mixture of proposals has \emph{discovered} both modes, but since the repulsion is not decreased, the proposals cannot concentrate around the mode. Plot~(d) shows \acro~with exponentially decayed repulsion $G_1=0.5$ (see Section \ref{sec_repulsion}), with the mixture proposal successfully approximating the target density. We denote this mixture as $\widetilde q^{(T)}(\x) = \frac{1}{N}\sum_{n=1}^N q_n^{(T)}(\x;\bmu_n^{(T)},\bSigma_n^{(T)})$, i.e., the mixture density composed of all the proposals at the last iteration of the algorithm. In all cases, we show also the marginal plots of $\widetilde q^{(T)}(\x)$ and $\widetilde \pi(\x)$.

\begin{figure*}[htp]  
\centering 
\begin{tabular}{@{}c@{}c@{}c@{}c@{}}
\includegraphics[width=0.5\columnwidth]{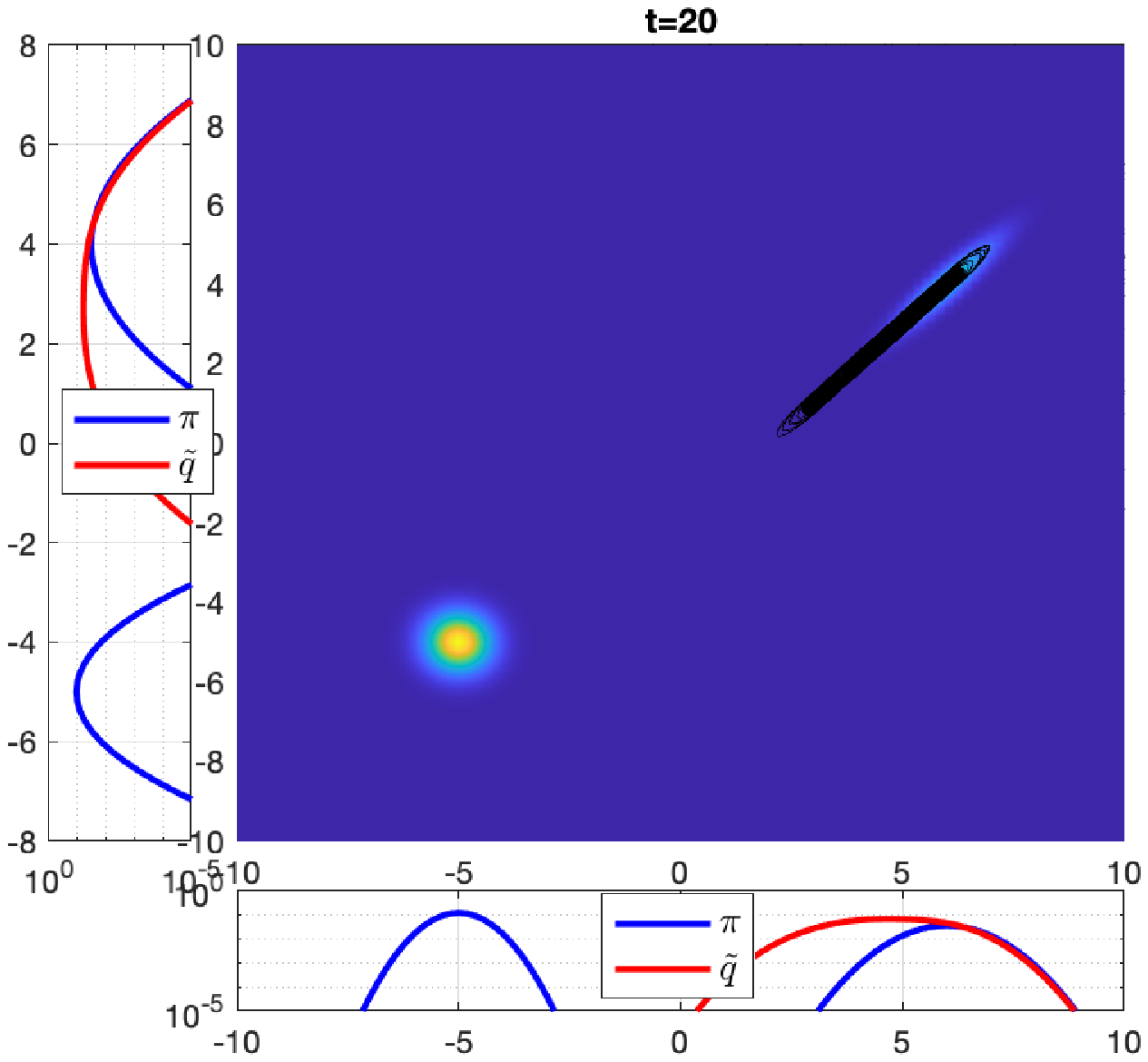} & \includegraphics[width=0.5\columnwidth]{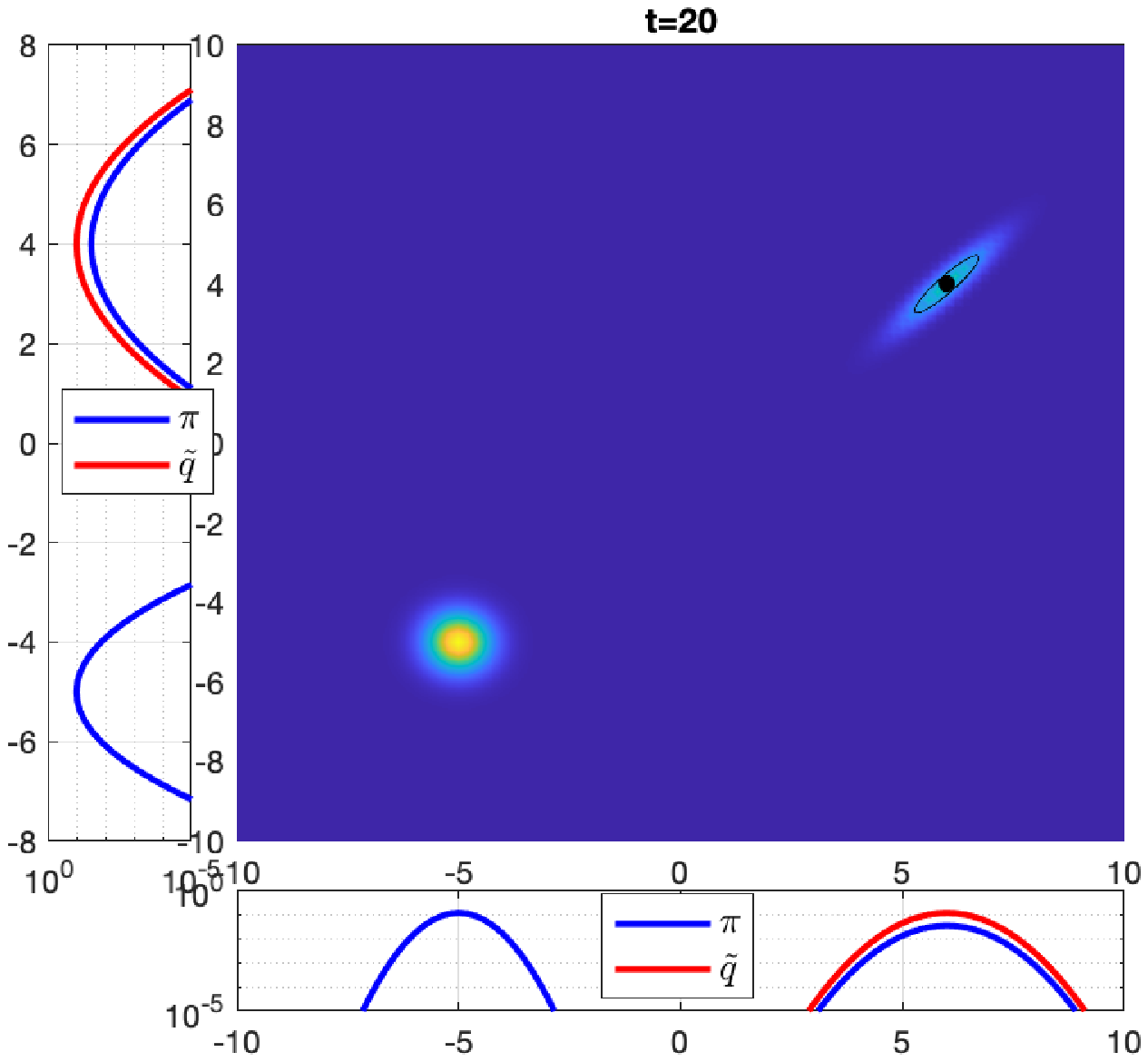} \\
\includegraphics[width=0.5\columnwidth]{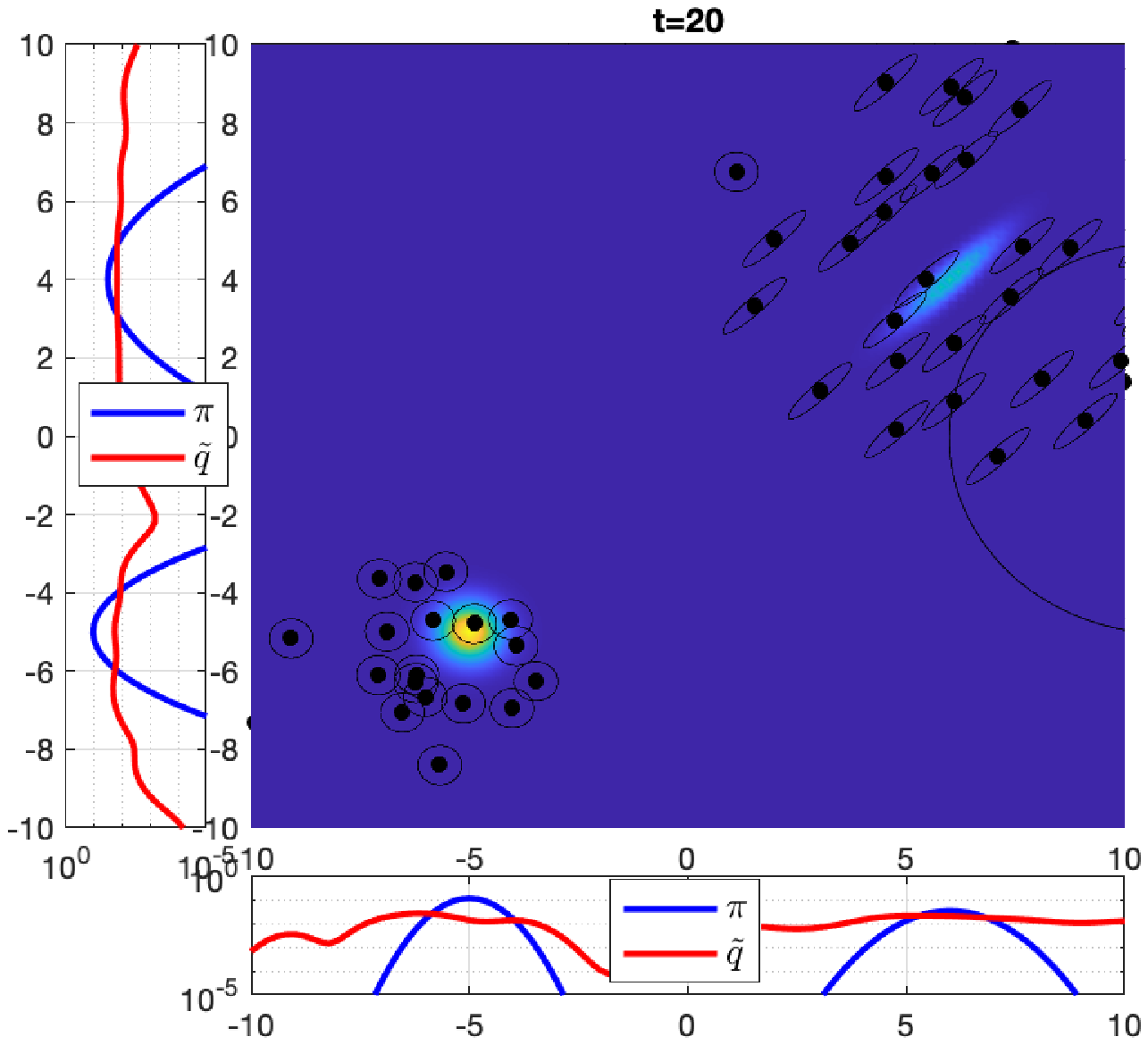} & \includegraphics[width=0.5\columnwidth]{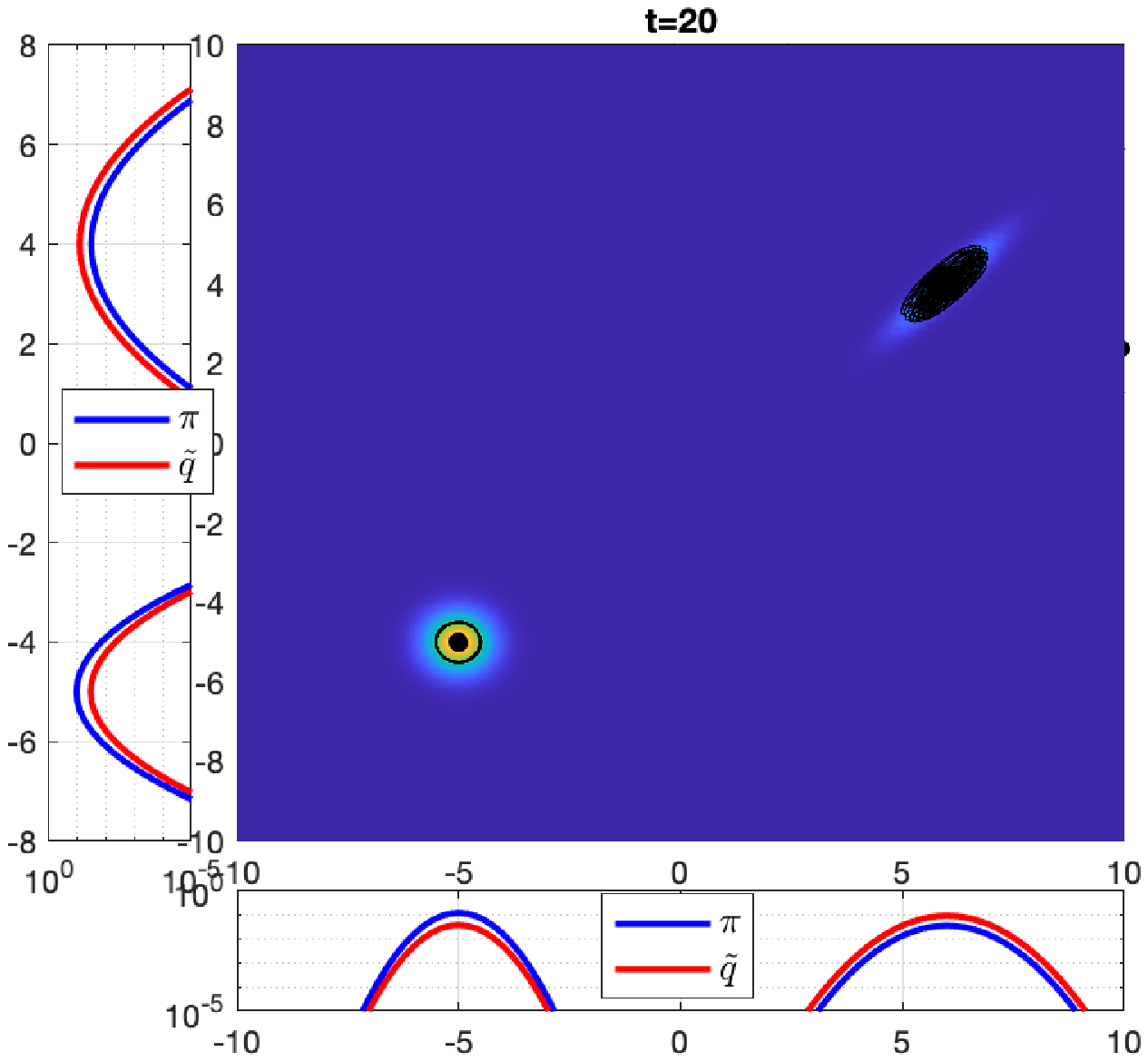} \\
\end{tabular}
\caption{\textbf{Toy example.} Final location parameters (black dots) and scale parameters (black ellipses) of the proposals at time $t=20$ for four ablated versions of \acro. Upper left: modified \acro~without pre-conditioning matrix in the gradient update (as in \cite{elvira2015gradient} with $\lambda=0.1$) and $G_t=0$ (no repulsion). Upper right: \acro~with $G_t=0$ (no repulsion). Bottom left: \acro~with constant repulsion $G_t=0.5$. Bottom right: \acro~with exponentially decayed repulsion $G_1=0.5$ (see Section \ref{sec_repulsion}).}
\label{fig_toy_2D}
\end{figure*}

We now extend the setting to $L=5$ components, with means ${\bf \gamma}_1=[-10, -10]^{\top}$, ${\bf \gamma}_2=[0, 16]^{\top}$, ${\bf \gamma}_3=[13, 8]^{\top}$, ${\bf \gamma}_4=[-9, 7]^{\top}$, ${\bf \gamma}_5=[14, -4]^{\top}$, and  covariance matrices $\bSigma_1=[5, \ 2; 2, \ 5]$, $\bSigma_2=[2, \ -1.3; -1.3, \ 2]$, $\bSigma_3=[2, \ 0.8; 0.8, \ 2]$, $\bSigma_4=[3, \ 1.2; 1.2, \ 0.5]$ and $\bSigma_5=[0.2, \ -0.1; -0.1, \ 0.2]$. 
This target is particularly challenging since it requires the algorithms to discover $5$  modes. We aim at estimating the first and second moments, and the normalizing constant, which are available in a closed form. 
%
%
The proposals are now randomly initialized in the square {$[-15,15]\times[-15,15]$}. 
%
%

Table \ref{table_exp1} shows the {RMSE} in the estimation of $Z$ and the first and second moments of the target in an ablation study of \acro. In particular, we test four versions of the algorithm, with/without preconditioning matrix in the update of the location parameters and with/without repulsion, i.e., the last column is the \acro~algorithm Table \ref{alg_gapis}. In the case without preconditioning matrix, we set $\gamma = 10^{-1}$.  In the case with repulsion, we use $G_1 = 0.05$ with exponential decay, otherwise we simply set $G_1 = 0$ to annihilate the repulsion effect. The MSE results are obtained over $100$ independent runs, with estimators using the weighted samples on the half last iterations. It can be seen that the worst results are obtained when no preconditioning and no repulsion are implemented, while the best results are obtained by the full \acro~algorithm.

\begin{table*} [h!]
\scriptsize{
\setlength{\tabcolsep}{2pt}
\def\marginwidth{1.5mm}    
\begin{center}
\begin{tabular}{|c||c|c|c||c|c|c||c|c|c||c|c|c| }                                 
\hline 
&  \multicolumn{3}{c }{No pre-cond./No repulsion} &  \multicolumn{3}{|c|}{No pre-cond./Repulsion} &  \multicolumn{3}{|c|}{Pre-cond./No repulsion}  & \multicolumn{3}{|c|}{Pre-cond./Repulsion}  \\
\cline{2-13}
\text{RMSE}  &  $\sigma=1$  & $\sigma=3$   & $\sigma=5$  &  $\sigma=1$  & $\sigma=3$   & $\sigma=5$  & $\sigma=1$  & $\sigma=3$   & $\sigma=5$ & $\sigma=1$  & $\sigma=3$   & $\sigma=5$   \\
\hline
\hline
$Z$ &   1.0108 &0.0339 &0.3152 &  0.0296 &0.0163 &0.0291  & 0.0224 &0.0128 &0.0192  &  \textbf{0.0096} &0.0168 &0.0264   \\
\hline
$\E_{\widetilde \pi}[{\bf X}]$ &  2.7567 &1.6222 &2.6798  & 0.7492 &0.7253 &0.5969 & 1.4756 &0.9557 &1.2745  & \textbf{0.7694} &0.9097 &1.5663   \\
\hline
$\E_{\widetilde \pi}[{\bf X}^2]$ &  2.5058 &1.7431 &2.3161  & 0.6521 &0.7439 &0.3422 &  1.6427 &0.8829 &1.7884 & \textbf{0.8137} &0.8895 &1.6063   \\
\hline
\end{tabular}
\end{center}
\caption{\textbf{Gaussians-mixture target in Section \ref{ex_mixture}.} {RMSE of the IS estimators. We run an ablation study of \acro~with/without preconditioning matrix in the update of the location parameters and with/without repulsion. In the case without preconditioning matrix, we set  $\gamma = 10^{-1}$.  In the case with repulsion,  $G_1 = 0.05$ with exponential decay. The MSE results are obtained over $100$ independent runs, with estimators using the weighted samples on the half last iterations.} 
\label{table_exp1}
}
}
\end{table*}

\subsection{{Generalized Gaussian mixtures}}
\label{ex_gg_mixture}

{
We consider a generalization of the previous study where the target is now a mixture of $L \geq 1$ bivariate generalized Gaussian pdfs as
\begin{equation}
(\forall \x \in \mathbb{R}^2) \quad \widetilde{\pi}({\bf x}) =  \sum_{\ell=1}^L \omega_\ell \mathcal{GG}(\x;\bnu_\ell,\bSigma_\ell,\eta_\ell),
\label{eq:mixGG}
\end{equation}
where, for every $\ell \in \{1,\ldots,L\}$, $\omega_\ell$ are the mixture weights, and $\bnu_\ell$, $\bSigma_\ell$, and $\eta_\ell$ are respectively the mean, scale, and shape parameters of each component of the mixture \cite{Gomez1998}, which is given by
\begin{equation}
 \mathcal{GG}(\x; \bnu_\ell,\bSigma_\ell,\eta_\ell) = \frac{d_x \Gamma(\frac{d_x}{2})}{ \pi^{\frac{d_x}{2}}
    \Gamma(1 + \frac{d_x}{2 \eta_\ell})2^{1 + \frac{d_x}{2 \eta_\ell}}}|\bSigma_\ell|^{-1/2} \exp\left(- \frac{1}{2} \left((\x - \bnu_\ell)^\top \bSigma_\ell^{-1} (\x - \bnu_\ell)\right)^{ \eta_\ell}\right).
 \label{gg_pdf}
\end{equation} 
The moments and a more standard parametrization are discussed in \ref{sec_append_parametrizations}. Generalized Gaussian mixture pdfs are popular in Bayesian inference approaches for image recovery \cite{Deledalle2018,FAN2009839,nguyen2022patch,CorbineauSPL}, as they constitute a rich and accurate model for texture components as well as for wavelet coefficient distributions. In this experiment, we set $L=5$ components, with means ${\bf \gamma}_1=[-10, -10]^{\top}$, ${\bf \gamma}_2=[0, 16]^{\top}$, ${\bf \gamma}_3=[13, 8]^{\top}$, ${\bf \gamma}_4=[-9, 7]^{\top}$, ${\bf \gamma}_5=[14, -4]^{\top}$, $\bSigma_\ell = I_{d_x}$, $\omega_\ell = 1/5$, and $\eta_{\ell} = \eta \in \{0.5,\;1,\;1.5\}$, for all $\ell=1,\ldots,5$. The location parameters of the proposals are now randomly initialized in the square {$[13,15]\times[-6,-8]$}. This initialization is particularly adversarial, since the location parameters of the proposals are very concentrated around only one of the five modes. We use the same parameters as in the previous example, except that GRAMIS now implements $G_t = 1$ to enforce exploration via the repulsion term. We compare GRAMIS with AMIS \cite{CORNUET12}, LR-PMC \cite{elvira2017improving}, and O-PMC \cite{elvira2022optimized} algorithms. We set the same parameters in all algorithms except in AMIS where, since it has only one proposal, we set $K=1000$ so all methods have the same number of target evaluations. Note that the generalized Gaussian pdf~\eqref{gg_pdf} is not differentiable at its mean as soon as $\nu < 1$ (e.g., $\nu = 0.5$ is Laplace distribution). In~\ref{app_smoothed}, we discuss how to circumvent this issue by building a smoothed approximation of the target, and explicit its gradients and Hessians (in the experiment we set $\delta = 10^{-5}$ for this smoothed version). Note that, for $\nu < 0.5$, the Hessian is not necessarily definite positive in all $\x \in \Real^{d_x}$. However, this does not create any instability in our implementation, thanks to the safe rule defined in \eqref{eq:covadapt1b} for the covariance adaptation.

The results are displayed in Table \ref{table_gg} in terms of RMSE in the estimation of $Z$ and the first and second moments. We can see that GRAMIS outperforms all competitors in all targets except in the second moment for the case with $\eta=0.5$ (Laplace distributions), where LR-PMC, O-PMC, and GRAMIS perform very similarly. This target has heavier tails and allows all algorithms to better explore the space compared to the cases with $\eta=1$ (Gaussian distributions) and $\eta=1.5$ (lighter tails). In this two last scenarios, all competitors get stuck in one or two modes while the repulsion of GRAMIS allows it to still discover the five modes. Finally, Table \ref{table_gg}  also includes the estimated $\chi^2\left(\widetilde \pi,\psi^{(T)}\right)$ divergence, where $\psi^{(T)}(\x) = \frac{1}{N}\sum_{n=1}^N q_n^{(T)}(\x)$ denotes the equally-weighted mixture of proposals in the last iteration. We averaged the estimated divergence over all the independent runs. We note also that the $\chi^2$ divergence is of particular interest in the case of importance sampling, since the variance of the estimator of the normalizing constant can be expressed as $\Var(\widehat Z) = \frac{Z^2}{N}\chi^2\left(\widetilde \pi,\psi^{(T)}\right)$ \cite{ryu2014adaptive,agapiou2017importance,miguez2017performance}. Thus, it is natural that in all cases, the ranking in performance is the same as for the RMSE of the estimator $\widehat Z$.

\begin{table*} [h!]
\scriptsize{
\setlength{\tabcolsep}{2pt}
\def\marginwidth{1.5mm}    
\begin{center}
{
\begin{tabular}{|c||c|c|c||c|c|c||c|c|c||c|c|c| }                                 
\hline 
&  \multicolumn{3}{c }{AMIS} &  \multicolumn{3}{|c|}{LR-PMC} &  \multicolumn{3}{|c|}{O-PMC}  & \multicolumn{3}{|c|}{GRAMIS}  \\
\cline{2-13}
\text{RMSE}  &  $\eta=0.5$  & $\eta=1$   & $\eta=1.5$  &  $\eta=0.5$  & $\eta=1$   & $\eta=1.5$  &  $\eta=0.5$  & $\eta=1$   & $\eta=1.5$  &  $\eta=0.5$  & $\eta=1$   & $\eta=1.5$  \\
\hline
\hline
$Z$ &  0.9692  & 0.9692      & 0.9693  &  0.0123 & 0.5145 &   0.6103 & 0.0091  & 0.6400 & 0.6261 &   \textbf{6.43}$\mathbf{\cdot10^{-4}}$ & \textbf{2.46}$\mathbf{\cdot10^{-6}}$ & \textbf{2.57}$\mathbf{\cdot10^{-3}}$ \\
\hline
$\E_{\widetilde \pi}[{\bf X}]$ &  56.08  & 56.07 & 56.13 &  1.2345  &    50.49
 &  55.17 &  0.6225 & 56.09 & 55.51 &  \textbf{0.0249}   & \textbf{1.49}$\mathbf{\cdot10^{-4}}$ &   \textbf{0.1097} \\
\hline
$\E_{\widetilde \pi}[{\bf X}^2]$ &   67.95  &   67.93   &  68.03&  1.119 & 55.57  &  65.80 & \textbf{1.0285}  &    67.89  & 66.62 &      1.2957 &  \textbf{0.0020}  &  \textbf{0.1875}  \\
\hline
\hline
$\chi^2\left(\widetilde \pi,\psi^{(T)}\right)$ &  980.12 &   980.07  &  980.08 & 30.61 &  556.74  & 616.8 &  7.836  &  639.93   &  630.27   &  \textbf{0.7781}  &  \textbf{0.0053}  & \textbf{4.3387}  \\
\hline
\end{tabular}
}
\end{center}
\caption{{\textbf{Mixture of generalized Gaussians in Section \ref{ex_gg_mixture}.} {RMSE of the IS estimators.    In the case of GRAMIS,  $G_1 = 1$ with exponential decay. The MSE results are obtained over $100$ independent runs, with estimators using the weighted samples on the half last iterations.}}
\label{table_gg}
}
}
\end{table*}

}

\subsection{Banana-shaped distribution}
\label{ex_banana}


We now consider a banana-shaped distribution \cite{haario1999adaptive,Haario2001}. The shape of this target makes it particularly challenging for sampling methods. The target is the pdf of a r.v. resulting from a transformation of a $d_x$-dimensional multivariate Gaussian r.v. $\bar{\X} \sim \mathcal{N}(\x;\textbf{0}_{d_x},\bSigma)$ with $\bSigma = \text{diag}(c^2,1,...,1)$. The transformed r.v. is $(X_j)_{1 \leq j \leq d_x}$ such that $X_j = \bar X_j$ for $j\in \{1,...,d_x\}\setminus 2$, and $X_2 = \bar X_2 - b(\bar X_1^2 - c^2)$, where we set $c=1$ and $b=3$ in our example.

First, we consider a toy example with $d_x=2$ so we can obtain intuitive plots. We set $T=100$, $N=50$, and $K=20$.  Figure \ref{fig_toy_banana_2D} shows the the target distribution, the final location parameters (black dots) and scale parameters (black ellipses) of the proposals at time $T=100$, and the samples of the last iteration (red dots). We consider the \acro~scheme with constant repulsion, with $G_t\in\{0.02, 0.01, 0.005, 0.001, 0.0001, 0 \}$. It can be seen that bigger values of $G_t$ yield effectively a mixture with well separated proposals. In this example, the proposals remain in practice static after a few iterations. When $G_t$ is smaller, the proposals tend to concentrate around the mode. In the extreme case with $G_t=0$ (i.e., without) repulsion, all proposals are effectively the same, which in practice coincides with the Laplace approximation \cite{shun1995laplace}.

We now perform comparisons with competitive algorithms, namely PMC using either global (GR) or local (LR) resampling \cite{elvira2017improving}, AMIS \cite{CORNUET12},  and O-PMC with LR \cite{elvira2022optimized}, for various dimension $d_x$. 
In AMIS, we set $N=1$, $K=500$ and $T=40$. The other algorithms set $N=50$, $K=20$, and $T=20$, so all algorithms have the same number of target evaluations. We measure the MSE of all algorithms in estimating $\E_{\widetilde \pi}[{\bf X}]$. 
%
   {In all cases, we select Gaussian proposal densities, with location parameters that are randomly initialized within the square $[-4,4]\times[-4,4]$.  
All algorithms are initialized with isotropic proposal covariances with $\sigma \in \{1, 3, 5\}$.}

{Table \ref{table_banana_gramis} shows the MSE of the proposed \acro~in the estimation of the target mean for $d_x\in \{5, 20, 50 \}$. 
In Fig.~\ref{fig_banana_dim}, we compare \acro, with {LR-PMC, GR-PMC} and O-PMC in a range of dimensions  $d_x \in \{2,5,10,15,20,30,40,50 \}$. The best performance is reached by \acro~in all dimensions, followed by the LR version of the O-PMC. We implement in this example a simple version of \acro~without repulsion, which simplifies the parameter tuning for different dimensions. In our \acro \; and in O-PMC, the MSE tends to  decrease when the dimensions grows, which can be explained by the strong structure of the banana-shaped target in high dimensions.}

Finally, Fig. \eqref{fig_banana_iter} shows the MSE in the estimation of $\E_{\widetilde \pi}[{\bf X}]$ versus the number of total iterations $T\in[10, 200]$ for the proposed \acro~method (with $\sigma=1$) using setting values of $G_{\text{rep}}\in \{0, \ 10^{-2} \}$. The estimators are built in all cases by using the last half of the total iterations. In the standard \acro~with $G_{\text{rep}} = 10^{-2}$, increasing the number of iterations improves the adaptation and thus the performance. However, if $G_{\text{rep}} = 0$ (i.e., no repulsion), running the algorithm for more iterations worsens the performance. This can be understood by seeing the last plot in Fig. \eqref{fig_toy_banana_2D}. In this case, all proposals tend to be the same, and thus the mixture does not represent well the whole target.

\begin{figure*}[htp]  
\centering 
\begin{tabular}{@{}c@{}c@{}c@{}c@{}}
\includegraphics[width=0.5\columnwidth]{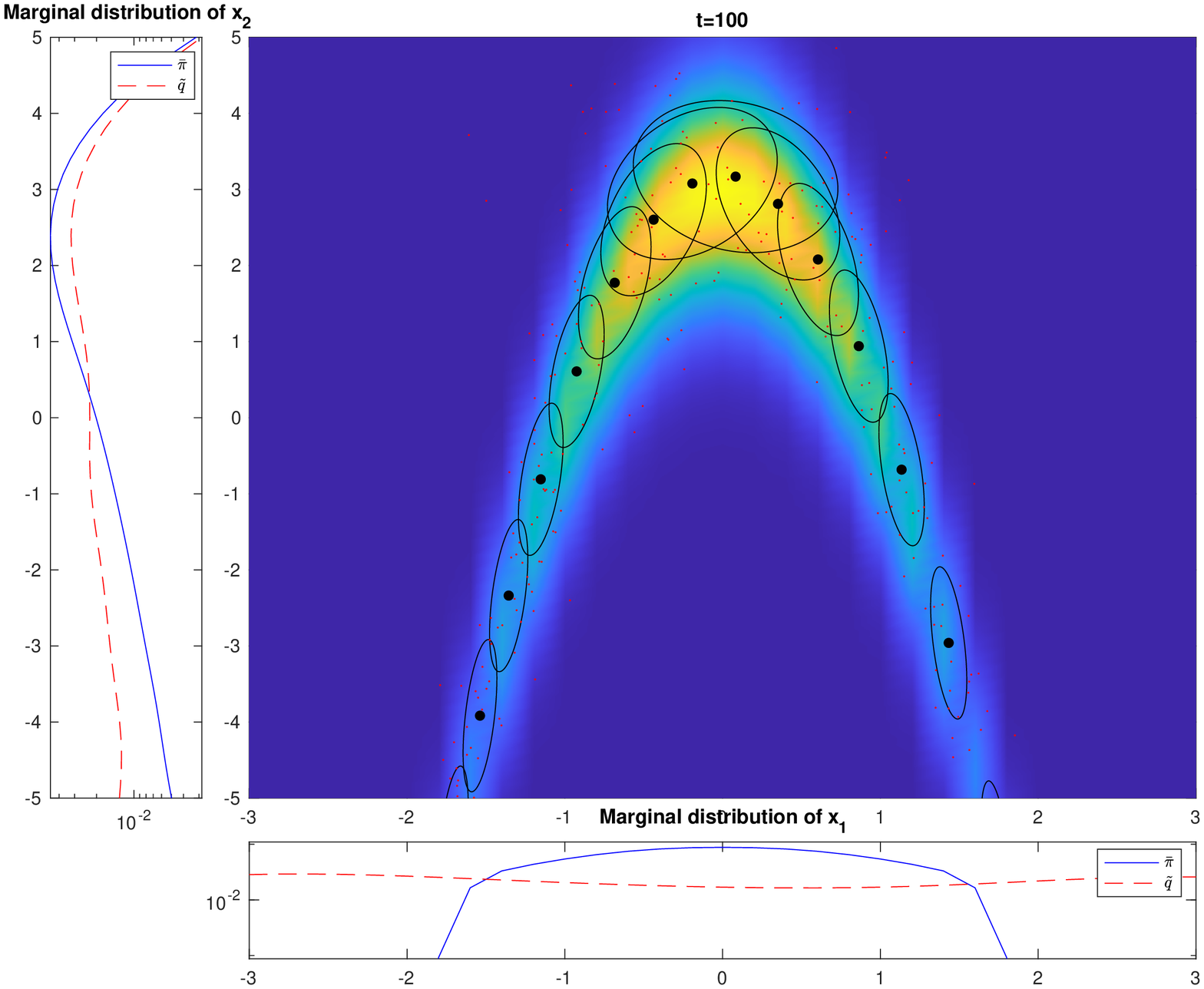} & \includegraphics[width=0.5\columnwidth]{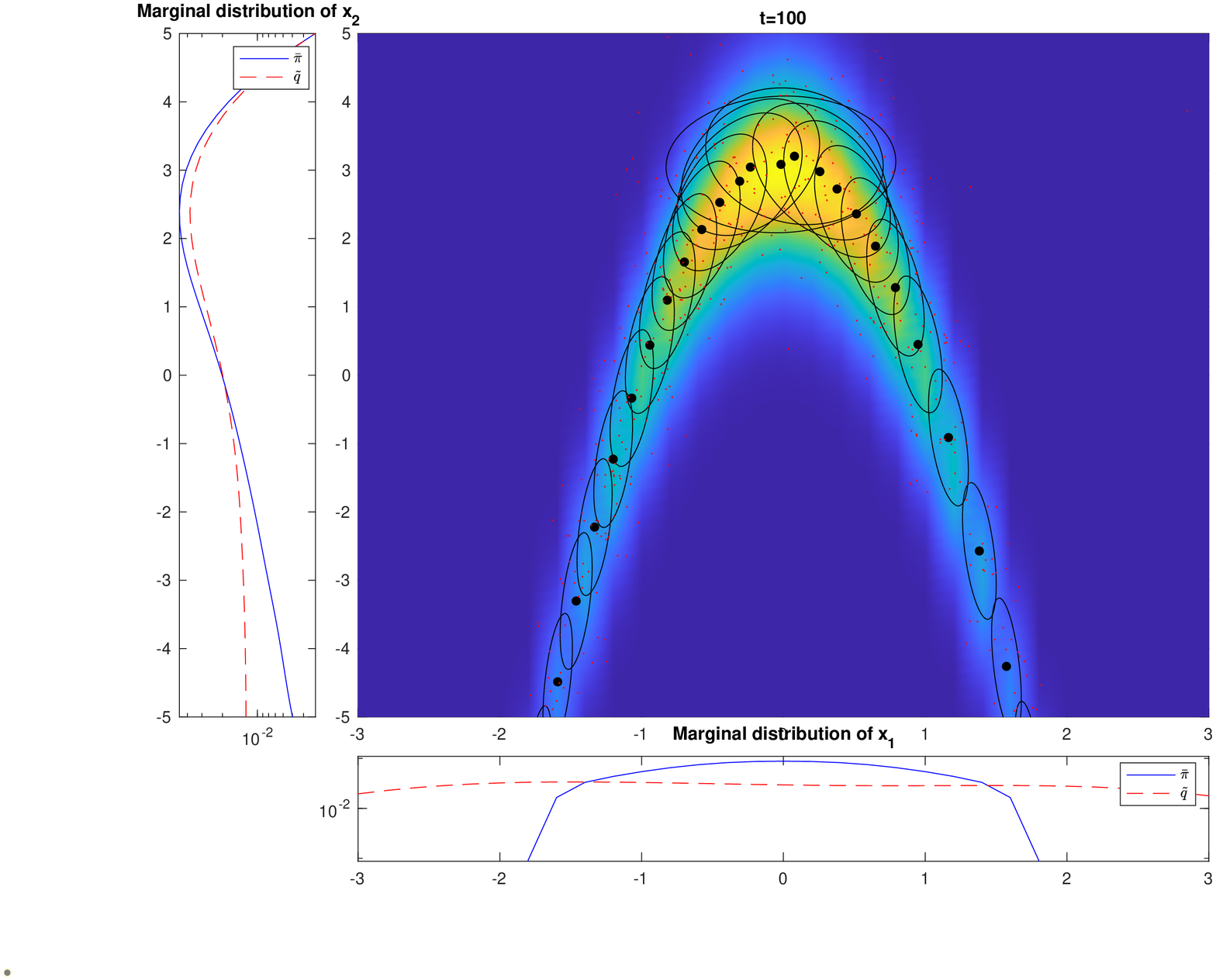} \\
\includegraphics[width=0.5\columnwidth]{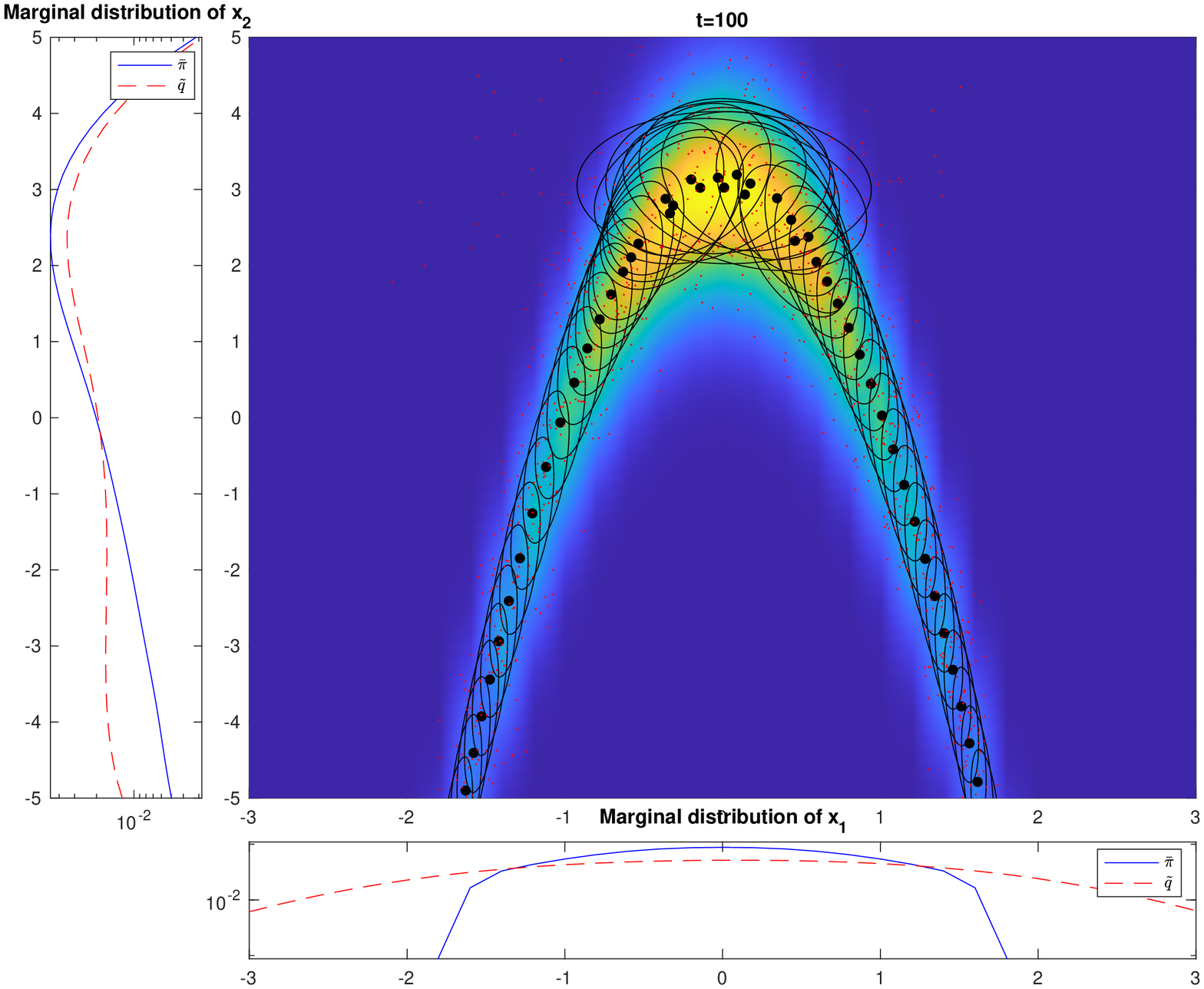} & \includegraphics[width=0.5\columnwidth]{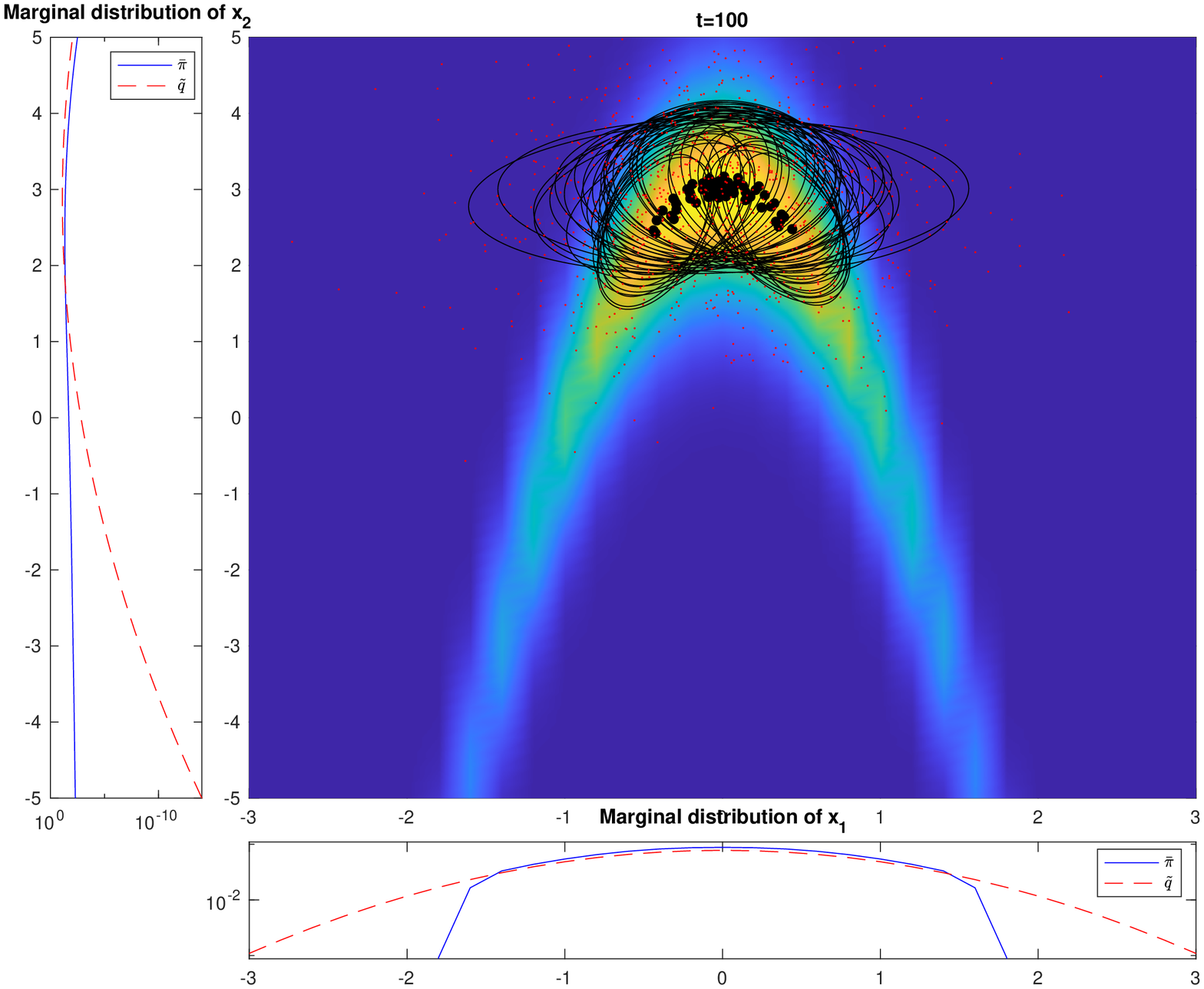} \\
\includegraphics[width=0.5\columnwidth]{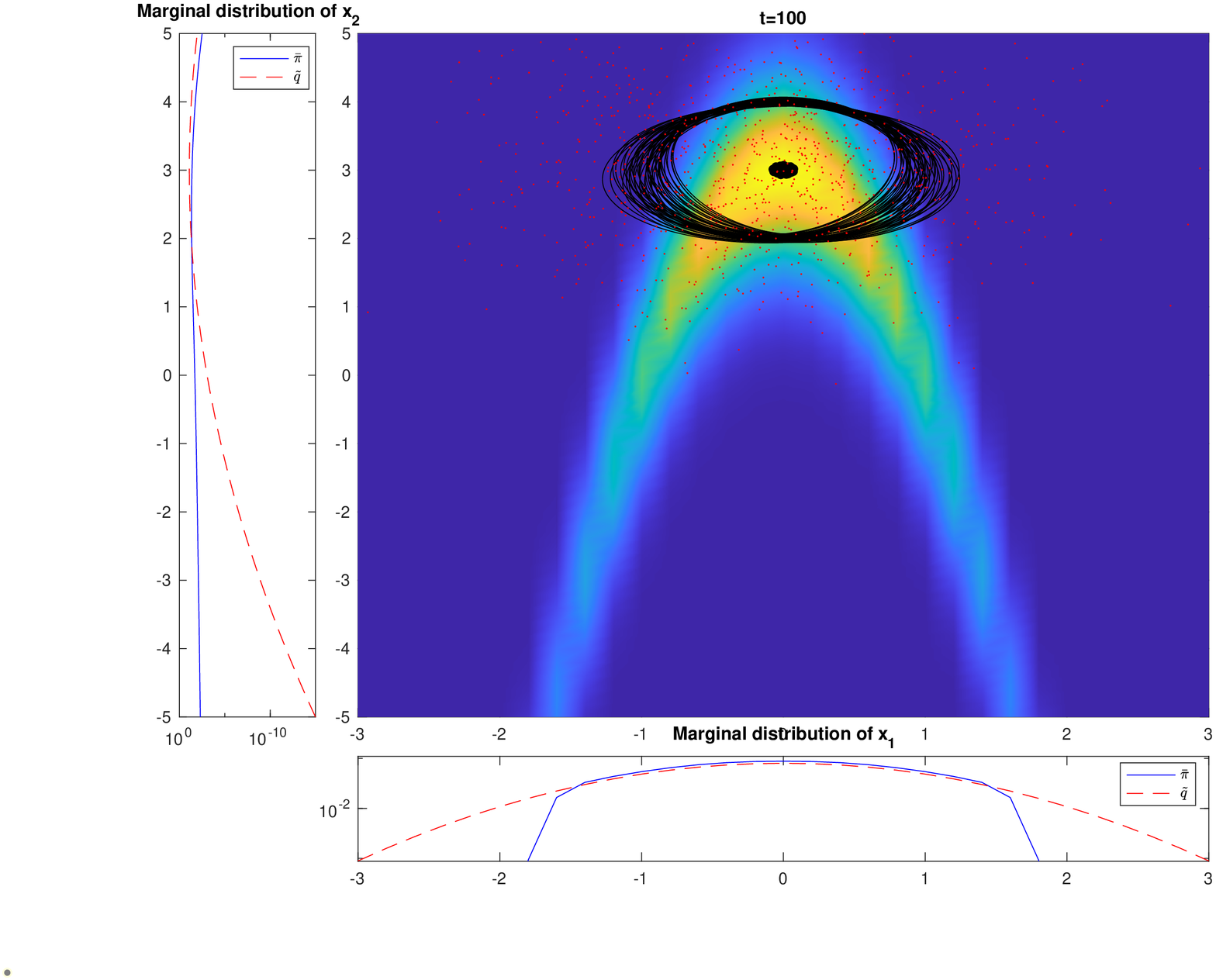} & \includegraphics[width=0.5\columnwidth]{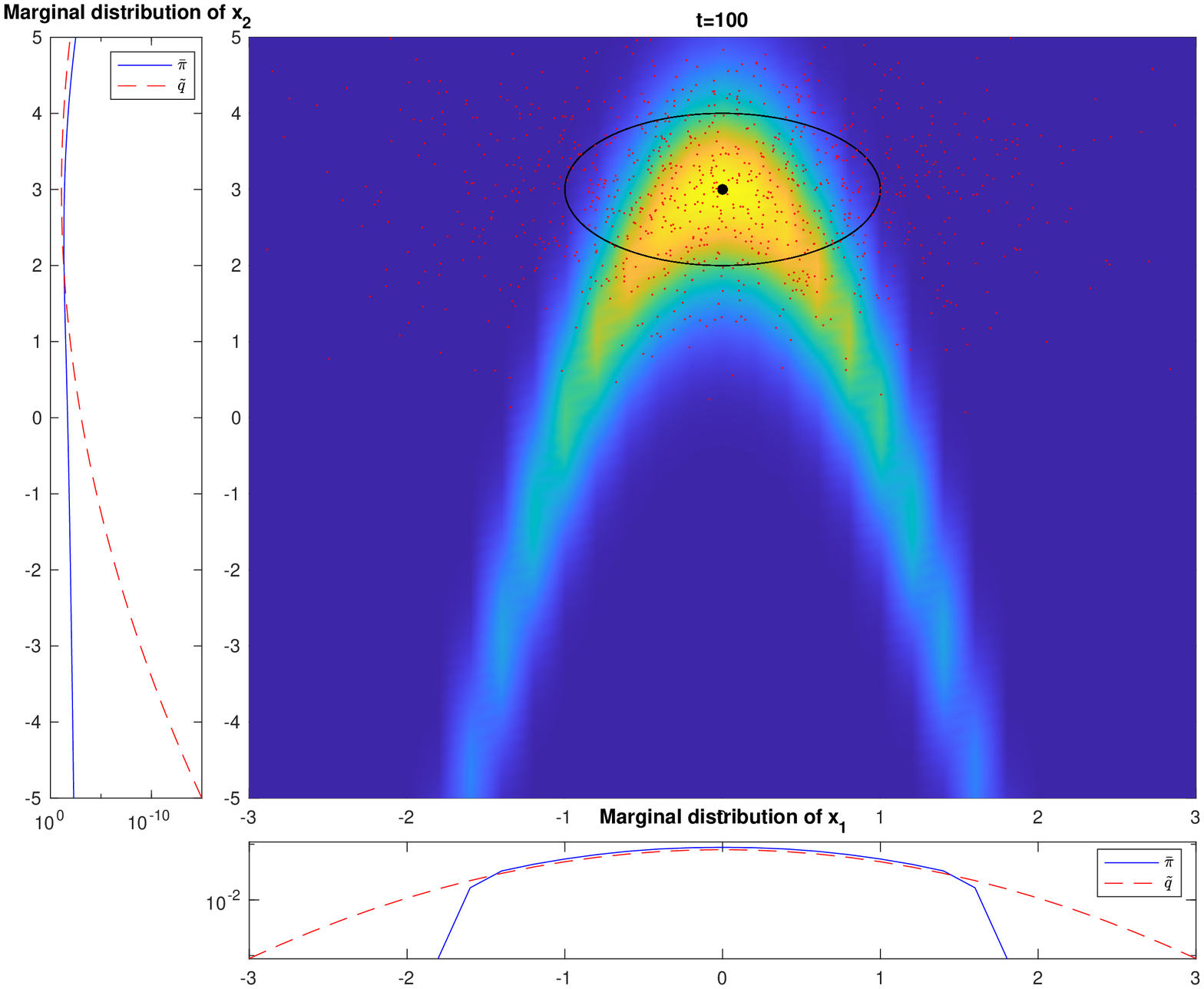} \\
\end{tabular}
\caption{  \textbf{Banana-shaped target in Section \ref{ex_banana}.} Final location parameters (black dots) and scale parameters (black ellipses) of the proposals at time $T=100$ for six ablated versions \acro~with constant repulsion. In order: $G_t\in\{0.02, 0.01, 0.005, 0.001, 0.0001, 0 \}$.}
\label{fig_toy_banana_2D}
\end{figure*}

\begin{figure}[h]
\centering
\includegraphics[width = 8cm]{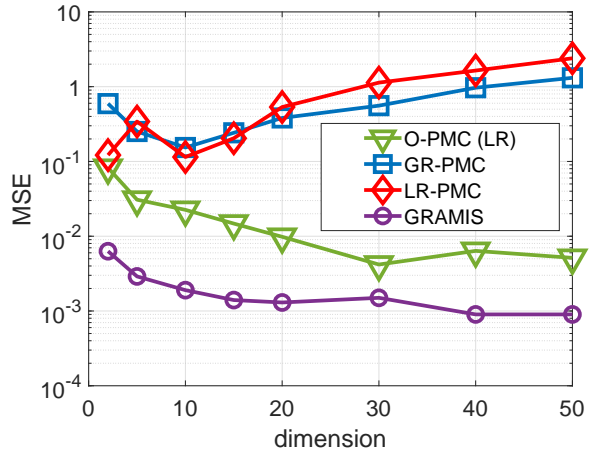}
\caption{\textbf{Banana-shaped target in Section \ref{ex_banana}.} MSE in the estimation of $\E_{\widetilde \pi}[{\bf X}]$ versus the dimension $d_x$, with GR-PMC, LR-PMC, O-PMC (using LR),  
and the proposed GRAMIS method (with $\sigma=1$).}
\label{fig_banana_dim}
\end{figure}

\begin{figure}[h]
\centering
\includegraphics[width = 8cm]{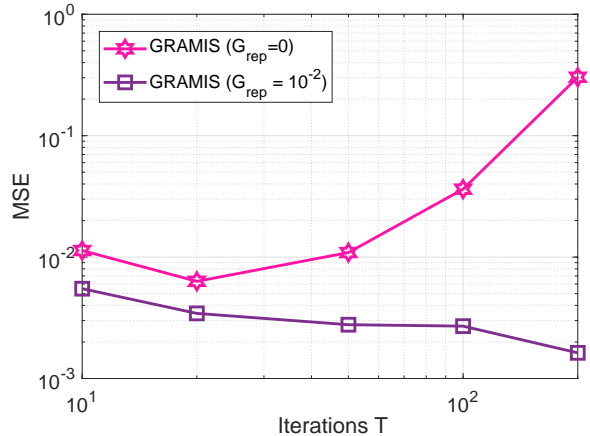}
\caption{\textbf{Banana-shaped target in Section \ref{ex_banana}.} MSE in the estimation of $\E_{\widetilde \pi}[{\bf X}]$ versus the number of iterations $T$ for the proposed GRAMIS method (with $\sigma=1$) setting $G_{\text{rep}}\in \{0, \ 10^{-2} \}$.}
\label{fig_banana_iter}
\end{figure}

\begin{table} [h!]
\scriptsize{
\setlength{\tabcolsep}{2pt}
\def\marginwidth{1.5mm}
\begin{center}
\begin{tabular}{|c|c|c|c|c|c|c| }                                 
\cline{2-7}
\multicolumn{1}{c|}{ } & GR-PMC  &  LR-PMC & AMIS  & LR-O-PMC & GAPIS & GRAMIS \\
\hline
\hline
$d_x = 5$ &  0.2515  & 0.3418  & 0.1758 & 0.0308 &  0.3007 &  \bf{0.0029} \\
\hline
$d_x = 20$ &  0.3818 & 0.5340 & 0.1901 & 0.0098 & 1.5299 &  \bf{0.0013} \\
\hline
$d_x = 50$ &  1.3134  & 2.3963 &  0.6074 & 0.0051 & 2.5524 & \bf{0.0009}\\
\hline                                                          
\end{tabular}
\end{center}
\caption{  \textbf{Banana-shaped target in Section \ref{ex_banana}.} MSE in the estimation of $\E_{\widetilde \pi}[{\bf X}]$ of the banana-shaped distribution for dimensions $d_x = 5$, $20$ and $50$. For all methods, we set the initial proposal variance to $\sigma = 1$. In all PMC-based methods, $(N,K,T) = (50,20,20)$ while $(N,K,T) = (1,500,40)$ for AMIS. 
}
\label{table_banana_gramis}
}
\end{table}

\section{Conclusion}
\label{sec_conclusion}

In this paper, we have proposed a new algorithm, called \acro, that iteratively adapts a set of proposals with the goal of improving the performance of the importance sampling estimators. The geometric information of the target is exploited by using the first-order and second-order information to improve the location and the scale parameters. A cooperation in the adaptation is allowed by introducing a repulsion term, which can be justified through the lens of Poisson fields. This repulsion becomes essential in multi-modal scenarios and also to represent target densities that, even if uni-modal, cannot be well approximated by standard uni-modal proposals. The \acro~algorithm exhibits good exploratory capabilities and a powerful representation of complicated target densities, leading in most cases to lower-variance estimators compared to other AIS algorithms. {As a future work, we plan to analyze the behaviour of the repulsion term in the proposals in GRAMIS algorithm in connection with a discretised Poisson field.}

\section*{Acknowledgement}

The work of V. E. is supported by the \emph{Agence Nationale de la Recherche} of France under PISCES (ANR-17-CE40-0031-01), the Leverhulme Research Fellowship (RF-2021-593), and by ARL/ARO under grants W911NF-20-1-0126 and W911NF-22-1-0235.

\appendix
{
\section{Gradient, Hessian, moments, and re-parametrization for the generalized Gaussian distribution}
\label{gg_appen}



\subsection{Smoothed approximation}
\label{app_smoothed}

We define the smoothed version of the multivariate generalized Gaussian distribution, as introduced in \cite{Marnissi2013} under the name \emph{generalized multivariate exponential power prior} (GMEP): 
\begin{equation}
(\forall \x \in \Real^{d_x}) \quad \mathcal{GMEP}(\x ; \bnu,\bSigma,\eta,\delta) = C |\bSigma|^{-1/2} \exp\left(- \frac{1}{2} \left((\x - \bnu)^\top \bSigma^{-1} (\x - \bnu) + \delta \right)^{ \eta}\right),
\label{eq:gmep}
\end{equation}
with $\delta>0$ and $C$ an appropriate normalization constant. The new parameter $\delta$ allows to smooth the distribution at $\x = \bnu$, so that \eqref{eq:gmep} is now twice differentiable on $\Real^{d_x}$. The original generalized Gaussian pdf is the case with $\delta = 0$. Note that the smoothing preserves the elliptical shape of the distribution. The practical application of such distribution to Bayesian inference for image recovery and remote sensing has been illustrated, for instance, in \cite{marnissiEUSIPCO,marnissi2018,Marnissi2018auxiliary}.

\subsection{Gradient expression}

We aim at calculating the gradient of $\log \widetilde{\pi}$, with $\widetilde{\pi}$ defined as in \eqref{eq:mixGG}. As aforementioned, we make use of the smoothed version \eqref{eq:gmep}, so that 
 \begin{equation}
    (\forall \x \in \Real^{d_x})\quad
    \nabla (\log \widetilde{\pi})(\x) = \frac{1}{\widetilde{\pi}(\x)} \nabla \widetilde{\pi}(\x), 
    \end{equation}
    with
\begin{align}
    (\forall \x \in \Real^{d_x})\quad
    \nabla \widetilde{\pi}(\x) & = \sum_{\ell=1}^L \omega_\ell \nabla \mathcal{GG}(\x ; \bnu_{\ell},\bSigma_{\ell},\eta_{\ell})\\
		& \approx \sum_{\ell=1}^L \omega_\ell \nabla \mathcal{GMEP}(\x ; \bnu_{\ell},\bSigma_{\ell},\eta_{\ell},\delta).
    \end{align}
Let us now explicit the gradient of \eqref{eq:gmep}. We introduce the short notation 
\begin{equation}
    (\forall \x \in \Real^{d_x})   (\forall \ell \in \{1 , \ldots,  L\}) \quad
    \theta_\ell(\x) = (\x - \bnu_\ell)^\top \bSigma_\ell^{-1} (\x - \bnu_\ell),
\end{equation}
and
\begin{equation}
    (\forall y\in \Real) (\forall \ell \in \{1, \ldots,  L\}) \quad g_\ell(y) = C_\ell \exp\left(- \frac{(y+\delta)^\eta}{2}\right),
		\label{eq:gell}
\end{equation}
so that
\begin{equation}
    (\forall \x \in \Real^{d_x}) \quad
  \mathcal{GMEP}(\x ; \bnu,\bSigma,\eta,\delta) = (g_\ell \circ \theta_\ell)(\x).
\end{equation}
Finally, using the chain rule:
\begin{equation}
    (\forall \x \in \Real^{d_x})   (\forall \ell \in \{1, \ldots,  L\}) \quad
 \nabla (g_\ell \circ \theta_\ell)(\x) = g'_\ell(\theta_\ell(\x)) \times \nabla \theta_\ell(\x),
\end{equation}
with the first derivative expression
\begin{equation}
  (\forall y \in \Real)   (\forall \ell \in \{1, \ldots,  L\}) \quad
g'_\ell(y) = - \frac{\eta_\ell}{2 } (y + \delta)^{\eta_\ell-1} g_\ell(y),
\end{equation}
and the gradient
\begin{equation}
        (\forall \x \in \Real^{d_x})   (\forall \ell \in \{1, \ldots,  L\}) \quad \nabla \theta_\ell(\x) = 2 \bSigma_\ell^{-1} (\x - \bnu_\ell).
\end{equation} 

\subsection{Hessian expression}

We proceed in a similar manner, using the smoothed version of the generalised Gaussian distribution. First,
\begin{equation}
    (\forall \x \in \Real^{d_x}) \quad 
    \nabla^2 (\log \widetilde{\pi})(\x) = - \frac{1}{\widetilde{\pi}(\x)^2}\nabla \widetilde{\pi}(\x)\nabla \widetilde{\pi}(\x)^\top  + \frac{1}{\widetilde{\pi}(\x)}\nabla^2 \widetilde{\pi}(\x),
\end{equation}
with
\begin{align}
    (\forall \x \in \Real^{d_x})\quad
    \nabla^2 \widetilde{\pi}(\x) & = \sum_{\ell\ell=1}^L \omega_\ell \nabla^2 \mathcal{GG}(\x ; \bnu_{\ell},\bSigma_{\ell},\eta_{\ell}),\\
		& \approx \sum_{\ell=1}^L \omega_\ell \nabla^2 \mathcal{GMEP}(\x ; \bnu_{\ell},\bSigma_{\ell},\eta_{\ell},\delta).
    \end{align}
Again, using the chain rule we obtain
\begin{multline}
    (\forall \x \in \Real^{d_x})   (\forall \ell \in \{1, \ldots,  L\}) \quad
 \nabla^2 (g_\ell \circ \theta_\ell)(\x) = g_\ell''(\theta_\ell(\x)) \times \nabla \theta_\ell(\x) \nabla \theta_\ell(\x)^\top \\
 + g'_\ell(\theta_\ell(\x)) \times \nabla^2 \theta_\ell(\x),
\end{multline}
where we have used the expressions 
\begin{equation}
        (\forall \x \in \Real^{d_x})   (\forall \ell \in \{1, \ldots,  L\}) \quad \nabla^2 \theta_\ell(\x) = 2 \bSigma_\ell^{-1},
\end{equation} 
and the second order derivative
\begin{equation}
 (\forall y \in \Real)   (\forall \ell \in \{1, \ldots,  L\}) \quad
 g_\ell''(y) = \left(- \frac{\eta_\ell(\eta_\ell-1)}{2} (y + \delta)^{\eta_\ell-2}
  + \frac{\eta_\ell^2}{4}(y + \delta)^{2 \eta_\ell - 2}\right) g_\ell(y).
\end{equation}

\subsection{Moments and re-parametrization}
\label{sec_append_parametrizations}
The generalized Gaussian distribution parametrized in \eqref{gg_pdf} has a mean $\E[X] = \bnu_\ell$ and a covariance $\text{Cov}[X] = \frac{2^{\frac{1}{\eta_\ell}}     
    \Gamma \left( \frac{d_x + 2}{2 \eta_\ell}\right)}{d_x\, \Gamma\left(\frac{d_x}{2 \eta_\ell}\right)}\bSigma_\ell$. 
		
A common parametrization for the scalar generalized Gaussian distribution is given by $\mathcal{GG}(x;\nu,\alpha,\beta) = \frac{\beta}{2\alpha\Gamma(1/\beta)} \; 
                          e^{-(|x-\mu|/\alpha)^\beta}$ instead of our choice $ \mathcal{GG}(x; \nu,\sigma,\eta) =  \frac{1}{ \Gamma(\frac{1}{2 \eta}) 2^{\frac{1}{2 \eta} } \sigma}  e^{- \frac{1}{2} \left(\frac{|x - \nu|}{\sigma}\right)^{ 2\eta}}$. It is easy to see by identification that the re-parametrization corresponds to $\beta = 2\eta$ and $\alpha = 2^{\frac{1}{2\eta}}\sigma$.
}

\end{document}